\documentclass[useAMS,usenatbib,onecolumn]{mn2e}

\usepackage{psfig}
\usepackage{amsmath}
\usepackage{marvosym}

\begin{document}

\title[Planet Searches]{Rounding up the wanderers:  optimizing coronagraphic
searches for extrasolar planets}

\author[Agol]{Eric Agol\\
Astronomy Department, University of Washington, Box 351580, Seattle, WA 98195
}

\maketitle
\begin{abstract}
I derive analytic scalings for coronagraphic imaging searches for extrasolar 
planets.  I compute the efficiency of detecting planets about any given star, 
and from this compute dimensionless distribution functions for the detected 
planets as a function of planet-star distance and distance to the host stars.
I find the following for blind
planet surveys: (1) the optimum wavelength is between 4000-5000~\AA\ for 
Earth-like planets and 4200-5800~\AA\ for Jovian planets; (2) between
21-32\% of the number of planets per decade of radius can be detected with
an optimized survey; (3) target stars should be 
ranked from greatest to least by their luminosity divided by distance to 
the sixth or eighth power, depending on the dominant source of noise
for the survey; (4) surveys targeting all main sequence stars will detect
$\sim$3 times as many planets as surveys only targeting G-type stars; and
(5) stellar populations with different metallicities should have exposure
times that vary with the cube of the metallicity.  
I apply these results to the current suite of proposed coronagraphic 
satellite telescopes, of which {\it TPF-C} is the most powerful, but a
much smaller telescope, {\it TOPS}, may have a significant chance of detecting
Earth-sized planets due to its small inner working angle and high throughput.  
The most significant
uncertainty in these results is the noise contribution of Exo-zodiacal light.
These results can be applied to designing coronagraphs, comparing proposed telescope 
designs, optimizing the observing strategies, determining the properties of detected
planet populations, and selecting target stars.
\end{abstract}

\section{Introduction}

Now that extrasolar planets have been discovered with various indirect
means \citep[radial velocity, transits and microlensing,][]{may95,kon03,bon04}, 
it is now a major goal of astronomy to directly detect extrasolar
planets.  Two means of imaging planets have been proposed:  optical 
coronagraphy to capture light from the star scattered by the planet
\citep{nis01,kuc02,kas03,bro03,sta05} 
and infrared interferometry to detect thermal emission from the 
planet\footnote{The recent detections of the eclipse of planet by their
host stars also constitute direct detection, but without knowing which photons 
are from the star and which are from the planet \citep{cha05,dem05}.}
\citep{leg96,fri04,kal05}.  Here I carry out a detailed analytic analysis of 
the prospects for planet detection using optical coronagraphy with
space-based telescopes.

A coronagraph enables planet detection by suppressing the light from the
central star and suppressing the wings of the point spread function.
The remaining stellar light is due to diffraction and scattering due to 
imperfections
in the optics, which in some designs can be suppressed further by differencing 
images taken
at different roll angles so that the speckles can be subtracted while
the planet remains.  Typical coronagraphic designs have an ``inner working
angle" within which the stellar PSF is too bright and the planet light
is suppressed by the coronagraph.  This sets the minimum angular scale at
which planets can be detected relative to their host stars.
Secondly, a coronagraph has 
a contrast between the wings of the PSF and the unobscured central
brightness of the star, which, in addition to flux from local zodiacal light
in our solar system and ``exo-zodiacal" emission due to scattering by
dust in the observed planetary system, sets the background at the position
of the planet.  These backgrounds set
the maximum planet-star separation at which the planet can be detected 
as the planet flux decreases as the separation increases.
Since the inner working
angle is proportional to the wavelength of light, at shorter wavelengths 
planets can be detected closer to their host stars (where they are brighter)
or around host stars at greater distances from Earth giving more stars
that may be probed.  However, the albedo and the stellar flux eventually 
decrease at short enough wavelengths, so at some point it does not pay to
have higher angular resolution.  The PSF contrast rises
as the inverse of the square of the wavelength \citep[due to the increase in
phase error for shorter wavelengths,][]{ang98}, which partly counteracts
the benefits of short wavelengths if the wings of the PSF dominate the
noise.  If zodiacal or exo-zodiacal light dominate the noise, then shorter
wavelengths means that the planet PSF will be more concentrated and thus
less affected by the zodi/exo-zodi surface brightness.
To compute the optimum wavelength
of observation requires a knowledge of the sample stars, coronagraph
properties, zodiacal and exo-zodicacal light, and survey parameters, as 
well as a model of the noise and
coronagraph response.  It is the goal of this paper to derive these quantities
in as simple a manner as possible to elucidate the optimization of
coronagraphic surveys for extrasolar planets.

As a measure of the success of a coronagraphic survey I use a straightforward
metric:  the total number of planets detected with specified properties.
This allows a quantitative comparison of various telescope 
designs, survey strategies, observation wavelength, and choice of stellar 
targets.  For the purposes of this study, I confine the analysis to blind,
single-visit surveys, i.e. for which there are no known planets and
each star is observed a single time.

Computations similar to those carried out here have been presented by 
\citet{bro05}, the primary difference being that Brown has focused on 
habitable terrestrial planets and his computations are numerical (using
Monte Carlo realizations of simulated surveys) which require great 
computational expense to explore parameter space for different surveys.  
The formulae presented here target detection of planets at any separation
from their host star and are analytic, allowing a rapid exploration
of parameter space and faster optimization of surveys.

After introducing my assumptions about the planets (\S \ref{ppp})
and coronagraphs (\S \ref{csa}), I compute the number of stars that can be surveyed 
(\S \ref{sss}), the efficiency of detecting planets about a given star (\S \ref{ddd}), and 
the planet and star distance distributions (\S \ref{ddd}).  I show how to choose the 
(dimensionless) survey volume (\S \ref{maxtv}), how to rank target stars
(\S \ref{rankstars}), and how to optimize the 
wavelength of observation (\S \ref{optlambda}).  I use my results to compare various
proposed coronagraphic imaging satellites (\S \ref{comptel}). I end with a discussion
of how to weight observing time by metallicity (\S \ref{metals}).
In \S \ref{conclusions} I discuss the results and outline topics for further 
study.  Two tables in the appendix summarize the notation used in the paper.

\section{Planet detection requirements} \label{detection}

To detect a planet with a coronagraphic telescope requires 
\begin{itemize}
\item sufficient signal-to-noise, 
$S/N \ge (S/N)_{det}$.
This should be chosen to be large enough that one is confident 
that a source is not due to a statistical fluctuation, so
$(S/N)_{det} \sim 5-10$ should be sufficient given that tens to
thousands of stars will be surveyed \citep[although higher 
signal-to-noise may be required to provide convincing evidence of
a planet,][]{gou04}.  
\item sufficient angular resolution,
$\theta =r \sin{\alpha} /D \ge \theta_{IWA}$, where $D$ is the distance to the star,
$\theta$ is the angular separation on the sky of the planet and star, $r$ is the planet-star
separation, $0\le \alpha \le \pi$ is the phase angle of the planet 
($\alpha=0$ at full phase), and $\theta_{IWA}$ is the 
inner-working-angle of the telescope within which a planet cannot be detected
near the star.
\end{itemize}
These two conditions lead to the differential number of planets a survey can detect,
\begin{equation} \label{eqn01}
{d N_{det} \over dr dR_p d\alpha dM dX dD} =
{1 \over 2} \sin{\alpha} \Omega_s D^2 {dn \over dM} {dp \over dX} {df \over dr dR_p} H(\theta-\theta_{IWA})
H(S/N - S/N_{det}),
\end{equation}
where $\Omega_s$ is the survey solid-angle,  $D$ is the distance to surveyed stars ($\Omega_s D^2$ is
the volume surveyed),
$dn/dM$ is the stellar mass function (I assume in this paper that host stars
are on the main-sequence), $dp/dX$ is the stellar
metallicity probability distribution ($X$=[Fe/H]), $df/(dr dR_p)$ (defined below) is the frequency
distribution of planets as a function of planet-star separation $r$ and
planet radius $R_p$, and $H(x)$ is the Heaviside step function.  I have assumed azimuthal symmetry
for the detectability of planets around a given star, so the angular detection limit just depends 
on $2\pi r^2\sin{\alpha}/(4\pi r^2)$.
I have assumed that the stellar metallicity distribution is independent of stellar mass,
hence the separate dependence on ${dn \over dM}$ and ${dp \over dX}$, and I have assumed
that the planet size distribution is independent of stellar properties.  The two step
functions enforce $\theta > \theta_{IWA}$ and $S/N > S/N_{det}$.
In the next section I specify
my assumptions about the planet properties used in integrating this equation.

\section{Planet parameter presumptions} \label{ppp}

In order to make a straightforward study of the dependence of coronagraphic
searches on the properties of the coronagraph and telescope, I 
make some simple assumptions about the planet distribution and physical 
properties of the reflecting planets.   
I assume that the distribution of planets is independent
of stellar spectral type, that the survey targets planets
of a particular size $R_p$, and that the planet frequency is
constant with the logarithm of radius,
\begin{equation} \label{planetdist}
{df \over d\ln{r} dR_p^\prime}= {f_{10}\over \ln{10}} \delta(R_p^\prime-R_p),
\end{equation}
where $f_{10}$ is average number of planets per decade of radius
and $\delta(x)$ is the Dirac delta function.
Since I am assuming single-visit observations of each star,
I can ignore the orbital parameters of each planet (such as
eccentricity) and simply treat the density of planets near each star as a power
law in radius once averaged over planet orbital elements.
This implies a spatial density of $n(r) = f_{10}/(4\pi r^3 \ln{10})$.
Now, if the planets are distributed between an inner radius, $r_{in}$,
and an outer radius, $r_{out}$, then the total fraction of stars with
planets is $f_p=f_{10} [\ln(r_{out}/r_{in})/\ln(10)]$.
I make this choice of planet distribution since the distribution of 
giant extrasolar planet periods is nearly proportional to 
$P^{-1}$ \citep{ste01,tab02,kuc04} which implies a constant number of 
planets per logarithmic radius interval, although one study suggests 
a shallower dependence \citep{lin03}.
We currently have no information on the distribution of terrestrial
planets, but the constant $\ln{r}$ distribution seems to be a sensible 
choice as it is not biased towards detecting planets at either larger 
or smaller radius.

I compute the planet brightness with several assumptions. First, I
assume that the brightness of
the planet scales as the inverse square of its distance from the
star.   Second, I assume that the geometric albedo has a fixed spectral
shape independent of planet-star separation.  Third, I assume that the 
phase function is independent of planet-star separation and wavelength.  
I assume that all of these
properties are independent of the planet mass, stellar spectral type,
age and metallicity.  
Then, for a single exposure of a star with a planet, the number of photons 
detected from the planet is 
\begin{equation} \label{eqn03}
Q_p \simeq Q_* p_{\lambda} \left({R_p \over r}\right)^2 \Phi(\alpha),
\end{equation}
where $p_{\lambda}$ is the geometric albedo (the fraction of the star's 
flux reflected at full phase) of the planet at the observed wavelength 
$\lambda$ , $\Phi(\alpha)$ is the
phase function (defined to be 1 at full phase when the star-planet-observer
angle $\alpha=0$), and $Q_*$ is the number of photons from the star.

Planetary phase functions are fairly complex, but given the generality 
of this paper, I use the phase function $\Phi(\alpha)=
\cos^4{(\alpha/2)}$.  I refer to this
as the ``quasi-Lambert" phase function as it approximates
the Lambert phase function, $\Phi(\alpha)=(\sin{\alpha}+(\pi-\alpha)
\cos{\alpha})/\pi$, which is the phase function of diffuse scattering. 
The quasi-Lambert phase function has a mathematically convenient form 
that allows for analytic solution of the planet detection efficiency.

\section{Coronagraph and survey assumptions} \label{csa}

I assume that the coronagraphic search is carried out
with direct imaging at a central wavelength $\lambda$ with a
fractional band-pass $\Delta\ln{\lambda} < 1$ with a mean throughput
of $\epsilon(\lambda)$.  The parameter $\epsilon$ includes all
inefficiencies due to, e.g., partial obscuration of the mirror by the 
secondary, partial 
reflectance, partial coronagraphic obscuration, quantum inefficiency,
observational overheads due to target acquisition and slewing, readout
time, and multiple exposures for covering multiple angular regions for 
non-axisymmetric PSFs.  I assume a circular telescope aperture
of radius $R_{tel}$ with an inner working angle 
\begin{equation} \label{eqn04}
\theta_{IWA}=\Theta_{IWA}
\lambda/(2 R_{tel}),
\end{equation} 
where $\Theta_{IWA}$ is a dimensionless
number, typically 1.5-5, which relates the inner working angle
and the ratio of the wavelength to the telescope diameter.

I assume that the telescope can achieve a contrast, $C(\lambda)$, which is 
constant in an annular region $\theta_{IWA}< \theta < \theta_{OWA}$,
where $\theta_{OWA}\gg \theta_{IWA}$ is the outer working angle.
The contrast is defined as the intensity ratio at a point in
the PSF to the intensity of the centre of an unocculted stellar
PSF \citep{kas03}.  Since the phase errors scale as $\lambda^{-1}$, I
assume
that the contrast ratio scales as $C(\lambda)=C_0(\lambda_0/\lambda)^2$,
where $\lambda_0$ is some reference wavelength \citep{mal95}.
This dependence results from gaussian errors in the phase \citep{ang98}.

I assume that the telescope carries out a blind survey of
nearby stars (i.e. not choosing stars that are known to have planets).
I assume a constant exposure time of each star, $T_{exp}$
(if each planet is observed multiple times at different roll angles
for speckle subtraction or covering the detection zone of the PSF, 
then $T_{exp}$ is the total time for
exposures which include the planet).
I assume that the survey has a fixed duration, $T_s$, 
which is typically of order several years, so that the total number
of stars surveyed is 
\begin{equation} \label{texpeqn}
N_s=T_s/T_{exp}
\end{equation}
(neglecting overheads, which
are likely small due to long exposure times).  Finally, I assume
that each star is observed at one epoch so that the planet is
at a fixed position.  I relax the constant exposure time assumption 
later.

\section{Signal-to-noise ratio} \label{snr}

The total signal-to-noise for an observation scales as 
\begin{equation} \label{sneqn}
S/N={Q_p \over \left(Q_Z + Q_{EZ} + Q_{PSF} + Q_B\right)^{1/2}},
\end{equation}
where $Q_p$ is the total number of detected photons from the planet, 
$Q_Z$ is the noise contribution from zodiacal light, $Q_{EZ}$ is
the noise contribution from exo-zodiacal light, $Q_{PSF}$ is the
noise contribution from the wings of the stellar PSF, and 
$Q_B$ is the noise contribution
from other sources (e.g. read-noise and dark current noise).  This
equation assumes the large-number photon limit applies, and
since I am computing the number of planets that can be detected
above a given signal-to-noise, I neglect the noise due to the 
planet (for planet characterization this should be included).
For the remainder of this paper I will ignore $Q_B$ as I expect
that detectors will be designed so that instrumental noise does
not dominate.

The total number of photons detected from the unobscured star is
\begin{equation} \label{eqn07}
Q_* \simeq \pi R_{tel}^2 \epsilon {L_\nu \over 4\pi D^2} \Delta\ln{\lambda} h^{-1} T_{exp}
\end{equation}
where $L_\nu$ is the star's specific luminosity (in erg s$^{-1}$ Hz$^{-1}$), 
$h$ is Planck's constant and the other quantities are defined above.
The noise due to zodiacal light is given by
\begin{equation} \label{eqn08}
Q_Z = {\pi R_{tel}^2 \epsilon L_{\odot,\nu} \Delta \ln{\lambda} h^{-1} T_{exp} 
\over 4 \pi (1 AU)^2} \tau_Z S_{fac} (\lambda/D_{tel})^2,
\end{equation}
where $\tau_Z$ (units of sr$^{-1}$) relates the solar flux to the zodiacal light surface
brightness in the direction of ecliptic longitude and latitude
$(\lambda_{ec},\beta_{ec})$ and $S_{fac}$ measures
the PSF sharpness in units of $\lambda/D_{tel}$ (e.g. for optimal PSF 
fitting with an Airy disc point-spread-function, $S_{fac}=2.1$ ).  
The noise due to exo-zodiacal light is given by
\begin{equation} \label{eqn09}
Q_{EZ} = {\pi R_{tel}^2 \epsilon L_\nu \Delta \ln{\lambda} h^{-1} T_{exp} 
\over 4 \pi (r \sin{\alpha})^2} \tau_{EZ} S_{fac}(\lambda/D_{tel})^2,
\end{equation}
where $\tau_{EZ}$ (units of sr$^{-1}$) relates the surface brightness of the exo-zodiacal
light at the position of the planet to the flux of the star at the
tangent point (integrated along the line of sight).
Finally, the noise due to the wings of the PSF is given by
\begin{equation}
Q_{PSF} = Q_* C(\lambda) S_{fac}.
\end{equation}

\section{Simple survey scalings} \label{sss}

Having outlined my various survey assumptions, I now derive some 
scaling parameters for coronagraphic surveys.  I first assume that
the noise is dominated either by $Q_Z$, $Q_{EZ}$, or $Q_{PSF}$;
in \S \ref{threenoise} I will combine these to determine
the scalings when all three sources of noise contribute.

\begin{figure}
\centerline{\psfig{file=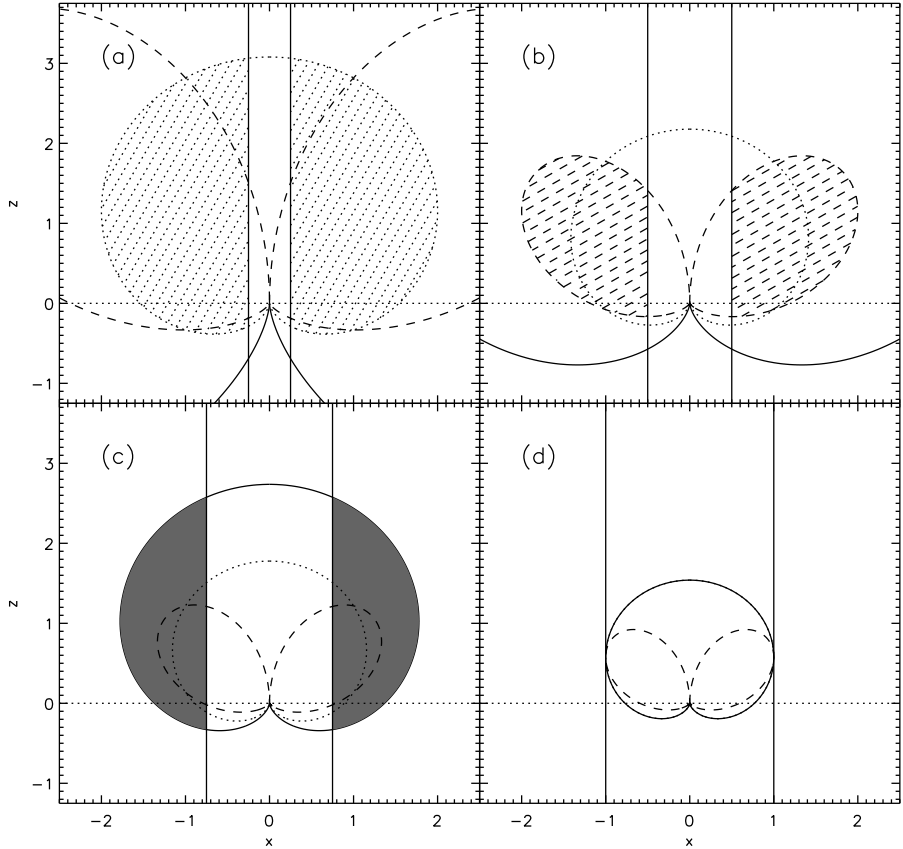,width=5in}} %Figure 1
\caption{Cylindrical cross section of the region where planets can be 
detected for stars at a distance $D$ from the Sun for (a) $D=0.25 
D_{max}$; (b) $D=0.5 D_{max}$; (b) $D=0.75 D_{max}$; (d) $D=D_{max}$.
Each panel is a 2-D cut through a 3-D region surrounding the star at 
the origin of each panel, where the units of the axes are
$\theta_{IWA} D_{max}$.
The observer is at $(x,z)=(0,-D\theta_{IWA}^{-1}D_{max}^{-1})$, and the inner working distance is
indicated with vertical solid lines.
The solid lines are for the Zodical dust noise limit, the dashed lines for
the Exo-zodiacal noise limit, and the dotted lines for the PSF-limited noise limit.
Note that for each noise limit, $D_{max}$ will
be different, so each panel compares the three noise limits at 
different physical distances.
The shaded regions are where a planet is detectable outside the
inner working angle and inside the flux limit (I have only shaded
one of the noise limits in each panel for clarity).  Since panel (d) is
at $D=D_{max}$, no planets can be detected. }
\label{fig01}
\end{figure}

There is a toroidal-shaped region surrounding the star within 
which the planet satisfies the signal-to-noise detection criterion
(Figure \ref{fig01}).  As the star becomes more distant the
signal-to-noise for planets at separation $r$ decreases
($\propto D^{[-1,-2,-1/2]}$ in the [Zodi, Exo-zodi, PSF]
limited cases), so the signal-to-noise detection region shrinks for a 
fixed exposure time.  The angular resolution limit corresponds to a 
cylinder surrounding the star with the axis along the line of sight
with a radius which grows as distance set by the inner working angle.
Inside the torus and outside the cylinder is the planet detection
zone.  At some distance, $D_{max}(T_{exp})$ these two limits meet
so that the torus is contained entirely within  the
cylinder - no planets can be detected for stars at $D>
D_{max}(T_{exp})$ for a particular set of planet, stellar and telescope 
parameters.  The distance $D_{max}$ sets
the scale for the volume of stars that can be surveyed of
a particular stellar type, so I next compute this for
each of the three noise limits.

\subsection{Zodiacal dominated noise} \label{zodinoise}

When $Q_Z \gg Q_{EZ}, Q_{PSF}$, to detect a planet requires
$Q_p/Q_Z^{1/2} \ge (S/N)_{det}$.  This implies
\begin{equation}\label{rmaxzodi}
r \le \left[{R_p^4 p_\lambda^2 \Phi(\alpha)^2 Q_*
{L_\nu \over L_{\odot,\nu}} \left({1 AU \over D}\right)^2 S_{fac}^{-1} \left({\lambda \over D_{tel}}\right)^{-2}
\tau_Z^{-1} (S/N)_{det}^{-2}}\right]^{1/4} = r_{2,Z}.
\end{equation}
The resolution limit requires
\begin{equation} \label{eqn12}
r \ge \Theta_{IWA} (\sin{\alpha})^{-1} \left({\lambda  \over D_{tel}}\right) D = r_1.
\end{equation}
Since detection requires $r_1 \le r \le r_{2,Z}$, the limiting distance 
$D_{max}$ is set
by $r_1=r_{2,Z}$ at $\alpha_{max}=60^\circ$ which is where the planet
has the largest angular separation from the star (this is where the 
flux-limit and resolution limit meet in panel d of Figure \ref{fig01}).
This gives
\begin{equation}\label{dmaxzodi}
D_{max}=D_Z \equiv D_{tel} \left[{\cal G} 
              {L_\nu^2 T_{exp}(1 AU)^2 \over \lambda^6 L_{\odot,\nu} \tau_Z \Theta_{IWA}^4 } \right]^{\delta}.
\end{equation}
where $\delta=1/8$ and
\begin{equation} \label{calgdef}
{\cal G} = {\textstyle \frac{1}{16}} R_p^4 p_\lambda^2 \Phi(\alpha_{max})^2 \sin^4{\alpha_{max}}
\epsilon \Delta\ln{\lambda}h^{-1} (S/N)_{det}^{-2} S_{fac}^{-1}.
\end{equation}

\subsection{Exo-zodiacal dominated noise}

The noise due to exo-zodiacal light is difficult to estimate as 
the zodiacal dust content of other old stellar systems is unknown, except for
a handful of infrared constraints with a sensitivity to dust $\sim$75
times as bright as our zodiacal dust cloud \citep{bei06}.  So, the
largest uncertainty in planet detection efficiency is due to
exo-zodiacal dust; until better constraints become available we will
have to rely on educated guesses to estimate the contribution of
this to planet detection noise.

For a planetary system with a dust cloud similar to that of the solar
system, the surface brightness declines steeply with distance from the
star, $\propto r^{-2.3}$, and varies significantly with the inclination
angle and azimuth of the disk due to flatness of zodiacal dust cloud
\citep{lei98}.
I will consider the case of largest noise: an edge-on disk.  In
this limit the equations simplify so that the exo-zodiacal surface
brightness is just a function of the angular separation of the planet
from the star.  The detection limit for signal-to-noise, $Q_p/Q_{EZ}^{1/2} 
\ge (S/N)_{det}$, translates into a similar limit on radius which is
\begin{equation}\label{ezsnr}
r \le \left[{R_p^4 p_\lambda^2 \Phi(\alpha)^2 Q_* D^{-2}
 \sin^2{\alpha}\ \tau_{EZ}^{-1} S_{fac}^{-1} D_{tel}^2 \lambda^{-2}
(S/N)_{det}^{-2} }\right]^{1/2} = r_{2,EZ}.
\end{equation}
This equation applies when $\tau_{EZ}$ is independent of $r$ (which is nearly
true for our zodiacal cloud).
The resolution limit is the same, so setting $r_1=r_{2,EZ}$ at
$\alpha_{max}$ (by coincidence, $\alpha_{max}=60^\circ$ in this
case as well) gives the maximum distance
\begin{equation} \label{dmaxez}
D_{max}=D_{EZ} \equiv D_{tel}\left[{\cal G} 
              {L_\nu T_{exp}\over \lambda^4 \tau_{EZ} \Theta_{IWA}^2 } \right]^{\delta},
\end{equation}
where $\delta = 1/6$ here.
For some stars with zodiacal dust that is more face-on, $D_{EZ}$
will be greater than this estimate.  However, given the uncertainties
in estimating $\tau_{EZ}$, a more complete calculation is unwarranted at
this point.

\subsection{PSF dominated noise}

The signal-to-noise limit in the PSF noise dominated case, 
$Q_p/Q_{PSF}^{1/2} \ge (S/N)_{det}$, leads to the requirement that
\begin{equation} \label{eqrmax}
r \le \left[Q_* C^{-1} S_{fac}^{-1} p_\lambda^2 R_p^4  
\Phi^2(\alpha) (S/N)_{det}^{-2}\right]^{1/4} = r_{2,PSF}
\end{equation}
Setting $r_1=r_{2,PSF}$ at $\alpha_{max}$ allows us to solve for
\begin{equation} \label{dmaxpsf}
D_{max}=D_{PSF} \equiv D_{tel}\left[{\cal G} 
              {L_\nu T_{exp}\over \lambda^2 C_0 \lambda_0^2 \Theta_{IWA}^4 } \right]^{\delta},
\end{equation}
where $\delta=1/6$ here.
With these expressions for $D_{max}$ in three limits, I
turn the planet detection expression into a dimensionless equation
in \S \ref{ddd}.

\subsection{Maximum number of stars} \label{maxnumstars}

I now compute the the maximum number of stars that may be surveyed of a particular
spectral type.
If all stars surveyed are of a specific spectral type, i.e. $dn/dM^\prime =
n_* \delta(M^\prime-M)$, then the total number
of stars surveyed is $N_s = \Omega_s  n_* D_s^3/3$, where
$n_*$ is the number density of stars of the given spectral type, assumed
to be constant, and $D_s<D_{max}$ is the limiting distance
of surveyed stars.  The total number of stars surveyed is also a function
of the survey duration, $N_s = T_s/T_{exp}$.

The maximum number of
stars that can be surveyed, $N_{max}$, can be found by equating these two
expressions for $N_s$ and setting $D_s=D_{max}=D_{[Z,EZ,PSF]}$.
I can solve for the maximum number of stars that can be surveyed in
the three different noise limits
\begin{equation} \label{numstar}
N_{max}=N_{[Z,EZ,PSF]}(T_s) = \left[{\textstyle\frac{1}{3}}\Omega_s n_* D_{[Z,EZ,PSF]}(T_s)^3 \right]^{1\over 1+3 \delta},
\end{equation}
where $D_{[Z,EZ,PSF]}(T_s)$ means that $T_{exp}$ is replaced by $T_s$ in equations
(\ref{dmaxzodi}),  (\ref{dmaxez}), or (\ref{dmaxpsf}).

Since $N_{max}$ sets the characteristic scale for the number of stars to survey 
for a particular set of telescope, stellar, and planet parameters as well as total
survey duration, it makes sense to scale the actual number
of stars surveyed as 
\begin{equation} \label{calneqn}
{\cal N} = N_s/N_{max},
\end{equation}
where ${\cal N} \la 1$.
In \S \ref{maxtv} I determine the value of $\cal N$ which maximizes the 
number of detected
planets and I estimate the total number of planets that can be detected, 
$N_{det}$, but it should be clear that $N_{det}$ scales with the maximum 
number of stars that can be 
surveyed, $N_{det} \propto N_{max}$, since the more stars that can be
surveyed, the more planets can be detected.   Although $N_{max}$ is
defined only for a specific spectral type, in \S \ref{rankstars} I generalize 
to surveys of stars with multiple spectral types.

\subsection{Summary} \label{threenoise}

When all three sources of noise contribute, the general equation
for $D_{max}(T_{exp})$ becomes:
\begin{equation} \label{eqndmax} % 8-14/06 notes
{D_{max}^8 \over D_Z^8} +{D_{max}^6 \over D_{EZ}^6} +{D_{max}^6 \over D_{PSF}^6} = 1,
\end{equation}
which is a quartic equation in $D_{max}^2$ with solution
\begin{eqnarray} \label{dmaxeqn}
D_{max}(T_{exp})&=&D_Z\left({G \over 2} -{V \over 4} +{1\over 2}\sqrt{{3V^2 \over 4}
-{V^3 \over 4 G}-G^2}\right)^{1/2} \cr
G&=& \left({V^2 \over 4}-{4 \over S}+{S \over 3}\right)^{1/2}\cr
S&=&{3 \over 2^{1/3}}\left(\sqrt{V^4 + 256/27} -V^2\right)^{1/3}\cr
V&=&{D_Z^6\left(D_{EZ}^6+D_{PSF}^6\right)\over D_{EZ}^6D_{PSF}^6}.
\end{eqnarray}

The ratio of the maximum distances in the three noise limits are
\begin{eqnarray} \label{caldeqn}
{\cal D}_1&=&{D_Z \over D_{EZ}} =(1 AU)^{1/4}\Theta_{IWA}^{-1/6}\lambda^{-1/12}\tau_{EZ}^{1/6}\tau_Z^{-1/8}L_\nu^{-1/24}{\cal G}^{-1/24}T_{exp}^{-1/24}L_{\nu,\odot}^{-1/8},\cr
{\cal D}_2&=&{D_Z \over D_{PSF}}=(1 AU)^{1/4}\Theta_{IWA}^{1/6}\lambda^{-5/12}\tau_Z^{-1/8}L_\nu^{1/12}{\cal G}^{-1/24}T_{exp}^{-1/24}C_0^{1/6}\lambda_0^{1/3}L_{\nu,\odot}^{-1/8}.
\end{eqnarray}
The weak dependence on $L_{\nu}$ in these equations means that different spectral
types will typically have similar noise properties.

Typically more than one source of noise dominates.  However,
the flux limits scale as $r_2 \propto D^{[-1,-2,-1/2]}$ in the 
Zodi, Exo-Zodi, and PSF dominated
cases, respectively, so if two  noise limits are comparable near
$D_{max}$, whichever grows fastest will dominate at smaller $D$.

When all three sources of noise contribute near $D_{max}$, the solution for
$N_{max}$ is 
\begin{equation} \label{nmaxeqn}
\left({N_{max} \over N_{Z}}\right)^{11/3}+
\left({N_{max}\over N_{EZ}}\right)^3 + \left({N_{max}\over N_{PSF}}\right)^3 = 1.
\end{equation}
This equation requires numerical solution of an eleventh order
polynomial for $N_{max}^{1/3}$; however, to an excellent approximation,
\begin{equation}\label{nmaxapprox}
N_{max}=\left(N_{Z}^{-10/3}+N_{EZ}^{-10/3}+N_{PSF}^{-10/3}\right)^{-3/10}.
\end{equation}

\section{Dimensionless distance distributions} \label{ddd}

The equations describing the detectable planets can be made
dimensionless by scaling the star distance, $D$,  as
$\eta=D/D_{max}\left(T_{exp}\right)$ and the planet-star separation, $r$, as
$\zeta=r/r_{max}$ where 
\begin{equation}\label{rmaxeqn}
r_{max} = {\theta_{IWA} D_{max}\over \sin{\alpha_{max}}}.
\end{equation}
Integrating the planet number density, $df/dr$ over the volume contained 
between the flux limits and inner working angle gives an expectation value 
for the number of planets detected about a star at a distance $\eta$ of
\begin{eqnarray} \label{ndetint}
{dN_{det} \over dM dX dD} &=& \Omega_s D^2 {dn \over dM} {dp \over dX}
2\pi \int_0^\pi \sin{\alpha} d\alpha \int_0^\infty dr {df \over dr} H(\theta-\theta_{IWA})
H(S/N - (S/N)_{det})\cr
&=& \Omega_s D^2 {dn \over dM} {dp \over dX}  E(\eta),
\end{eqnarray}
where $E(\eta)$ is the expectation value of the number of planets of radius $R_p$
detected for an observation of a single star with mass $M$, metallicity $X$, and
distance $D=\eta D_{max}(T_{exp})$.

\subsection{Expectation value as a function of distance} \label{expsection}

\begin{figure}
\centerline{\psfig{file=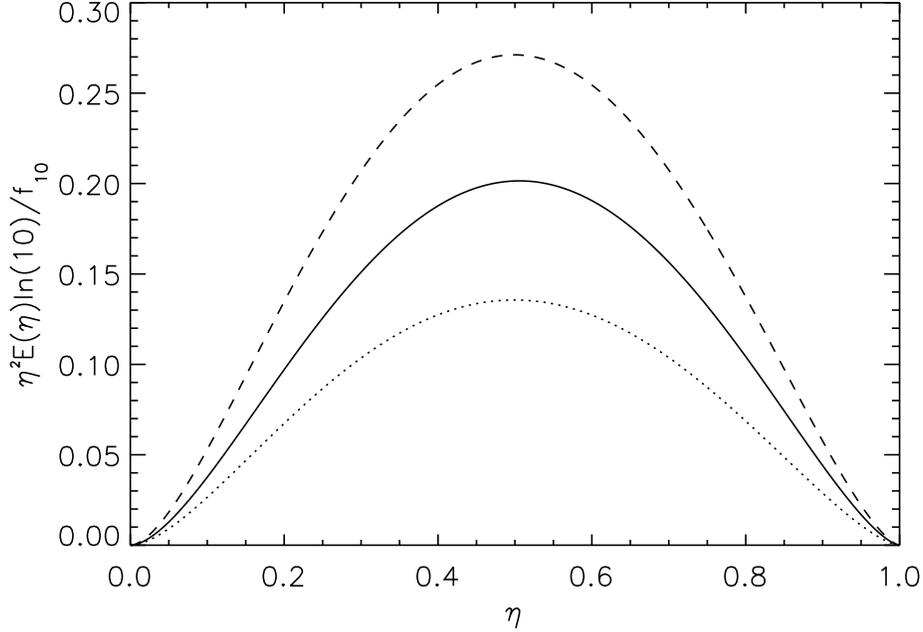,width=5in}} %Figure 2
\caption{The expectation value for the number of planets detected as a function
of distance for Zodi-limited (solid), Exo-zodi limited (dashed), and 
PSF-limited (dotted) surveys multiplied by the volume factor $4\pi\eta^2$ as
a function of $\eta=D/D_{max}$.}
\label{fig02}
\end{figure}

In the limit that one source of noise dominates at all distances, I can
compute the integral of $E(\eta)$ over $\alpha$ and $r$, finding
\begin{eqnarray} \label{eqleqn}
E(\eta)&=& \phi f_{10}(\ln{10})^{-1} \left[3 \ln{\left(x_+/x_-\right)}-(x_+-x_-)\right], \cr  % 7/1/06 notes
x_{\pm}&=& (1+\Delta)\left[1\pm \left(2\Delta^{-1}-1\right)^{1/2}\right], \cr
\Delta &=& \left(1+{3s \over 2 \Gamma} + {3 \Gamma \over 2}\right)^{1/2}, \cr
\Gamma &=& \left[s\left(1+\sqrt{1-s}\right)\right]^{1/3},
\end{eqnarray}
for $0 \le \eta \le 1$, where $s=\eta^{1/(2\delta)}$
and $\phi=1/2$ applies in the PSF and zodi limits, while
$\phi=1$ in the exo-zodi limit.  This can be approximated accurately by 
$\tilde E(\eta) =\phi f_{10}{13 \over 6\ln{10}} s^{-1/6}(1-s^{1/3})^{3/2}$. % for $s \ga 0.007216$.
In carrying out this integral, I have neglected the cutoffs in the planet 
distribution at $r_{out}$ and $r_{in}$ which is appropriate if
the peak of the detected planet distribution is well within these radii.
If $df/d\ln{r}$ is a power law with radius rather than a constant, then
$E(\eta)$ can be expressed in terms of Hypergeometric functions; however,
for the reasons discussed above I stick to using a uniform $\ln{r}$ distribution.
These three functions are plotted in Figure \ref{fig02}, which both show a 
rise at small $\eta$ as a larger volume is probed and fall at $\eta <1$ 
as the detection threshold is reached.  %The thermal phase function

In the limit ${\cal N}=1$, the total planet fraction can be computed analytically
for the three noise limits.
The fraction of surveyed stars with detected planets is
\begin{equation} \label{deteff}
{N_{det} \over N_{max}} = 3\int_0^1 \eta^2 E(\eta) = f_{10}
\begin{cases} 5 \pi 2^{-7/2} 3^{-5/4} \left[\ln(10)\right]^{-1}\simeq 0.153 & \text{Zodi-limit} \cr
{64 \over 135} \left[\ln(10)\right]^{-1}\simeq 0.206  & \text{Exo-zodi limit}\cr
{32 \over 135} \left[\ln(10)\right]^{-1}\simeq 0.103  & \text{PSF-limit}
\end{cases}
\end{equation}
So, depending on which noise-limit dominates 10-20\% of planets per decade of
radius will be detected around the stars that are surveyed.
The detection inefficiency is due to both the spatial resolution limit and the flux 
limit which allow detection of planets within a limited range of radii, which I
compute in the next section.

In the general case when all sources of noise contribute I have to integrate
the dimensionless detection equations numerically.  Defining
\begin{eqnarray}  \label{eqn30}
\eta_Z &=& {D_{max} \over D_Z}, \cr
\eta_{EZ} &=& {D_{max} \over D_{EZ}}, \cr
\eta_{PSF} &=& {D_{max} \over D_{PSF}},
\end{eqnarray}
where $0\le \eta_Z,\eta_{EZ},\eta_{PSF} \le 1$ since $D_{max} \le D_Z, D_{EZ}, D_{PSF}$,
then,
\begin{equation}\label{etaeqn}
\eta_Z^8 + \eta_{EZ}^6 + \eta_{PSF}^6 = 1,
\end{equation}
which follows from equation (\ref{eqndmax}).
The equations for  $r_1$ and $r_2$ become
\begin{eqnarray} \label{eqn32}
{r_1 \over r_{max}} &=& \eta {\sin{\alpha_{max}} \over \sin{\alpha}},\cr
{r_2 \over r_{max}} &=& \left[{\left(\eta_{EZ}^{12} + 4\eta^{-4}(\eta_Z^8 + \eta_{PSF}^6 \eta^{-2})
\left({\Phi^2 \sin^4{\alpha} \over \Phi_{max}^2 \sin^4{\alpha_{max}}}\right)\right)^{1/2}
-\eta_{EZ}^6 \over 2 \left(\eta_Z^8+\eta^{-2}\eta_{PSF}^6\right) {\sin^2{\alpha} \over 
\sin^2{\alpha_{max}}}}\right]^{1/2},
\end{eqnarray}
which is just a way of rewriting the resolution and signal-to-noise limits on radius.
Using equation \ref{etaeqn}, I can eliminate $\eta_{Z}$ so that $r_2/r_{max}$ is just
a function of $\eta$, $\eta_{EZ}$, $\eta_{PSF}$ and $\alpha$.
With these definitions, the planet expectation value for a star at distance $\eta$
is given by
\begin{equation} \label{eqn33}
E(\eta,\eta_{EZ},\eta_{PSF}) = {f_{10} \over 2\ln{10}}
\int_{x_-/2-1}^{x_+/2-1} d(\cos{\alpha}) \ln{\left({r_2 \over r_1}\right)},
\end{equation}
where $x_\pm=4\cos^2{\alpha_{\pm}\over 2}$ are given by equation (\ref{eqleqn}), but with 
$s=\eta^3\left[1-\eta_Z^8(1-\eta^2)\right]^{1/2}$ ($\alpha_{\pm}$ are the angles
where the resolution limit crosses the flux limit).
I have not found an analytic solution to this integral, so I integrate it numerically below.
The shape is qualitatively the same as in the three noise limits since all three
have very similar shapes.

\subsection{Radius distribution of planets}

\begin{figure}
\centerline{\psfig{file=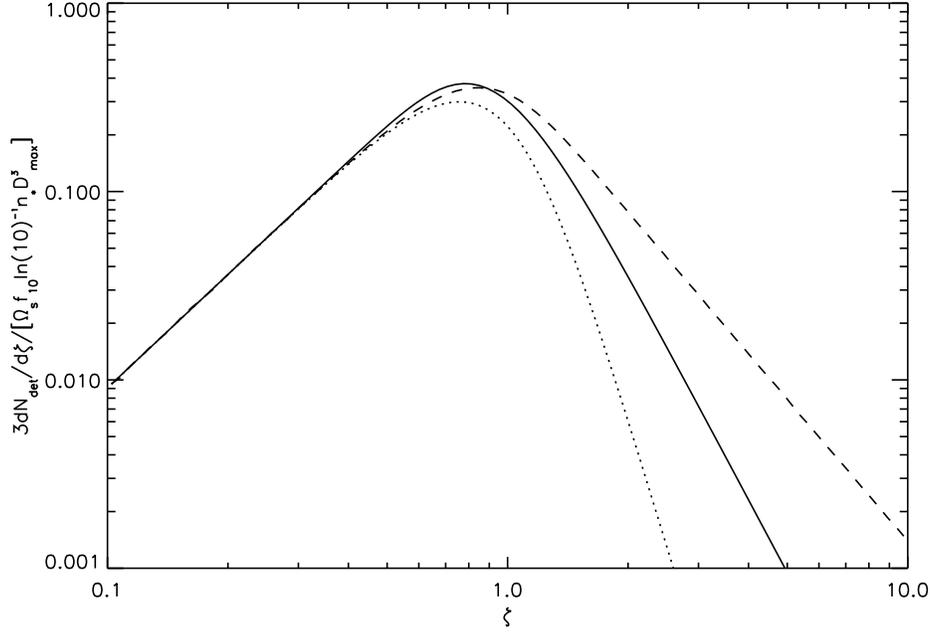,width=5in}} %Figure 3
\caption{The expected distribution of the detected planet-star separations
$\zeta=r\sin{\alpha_{max}}/[\theta_{IWA}D_{\max}]$ for the Zodi-limit
(solid line), Exo-zodi limit (dashed line), and PSF-limit (dotted line)
for surveys of a single stellar type, constant exposure time per star,
and ${\cal N}=1$.}
\label{fig03}
\end{figure}

In the limit ${\cal N} =1$, the radial distribution of detected planets can 
be computed analytically in the Zodi and PSF noise limited cases,
assuming that all stars observed are of a particular spectral
type (e.g. G stars).  For the Zodi limit,
\begin{equation}
{d N_{det} \over d\zeta} = {\Omega_s n_* D_{max}^3 f_{10} \over 3\ln{10}}
\left[{\zeta^4(1-\zeta^4)(1+9\zeta^4) \over (1+3\zeta^4)^3} +
{\zeta^2 \over 2\sqrt{3}} \cos^{-1} \left({3 \zeta^4-1 \over
3\zeta^4+1}\right)\right].
\end{equation}
The Exo-zodi case can be computed analytically in terms of
Hypergeometric functions, but an excellent analytic approximation is
\begin{equation}
{d N_{det} \over d\zeta} = {\pi^2\Omega_s n_* D_{max}^3 f_{10} \over 4\ln{10}}
\left({1 \over \zeta^{-2/g} + \zeta^{7/g}}\right)^g,
\end{equation}
where $g=1.9$ provides a good fit to the numerical curve.
For the PSF limit,
\begin{eqnarray} \label{eqn35}
{d N_{det} \over d\zeta} &=& {\Omega_s n_*D_{max}^3 f_{10} \over 2\zeta \ln{10}} 
\left[\frac{w^6}{\zeta^6} \left(\frac{1}{3} - \frac{3 w}{7}\right)+
{w^2 \over 6}\left(1-{3w\over 2}\right)\left(1-6w+{9 w^2 \over 2}\right)
+{\zeta^3 \over \sqrt{27}}\cos^{-1}\left({3w\over2}-1\right)\right],\cr
w&=&\zeta^3\left[\Sigma-\Sigma^{-1}\right],\cr
\Sigma &=& \zeta^{-1}\left(2+\sqrt{\zeta^6+4}\right)^{1/3}.
\end{eqnarray}
These
distributions are plotted in Figure \ref{fig03}.  All three distributions
peak near $\zeta=1$ since the largest number of stars surveyed are near 
$D_{max}$ where the sensitivity is concentrated near $\zeta=1$ since this
is near where the resolution and flux limits meet.
Each of these curves
approaches the same value at small $\zeta$ since the planets closest
to the star are limited by resolution, which is independent of the
phase function, scaling as $\propto \zeta^2$.  At large $\zeta$, planet 
detection is flux-limited which scales as $r_2 \propto D^{-b}$
where $b=[1,2,1/2]$ in the Zodi, Exo-zodi, and PSF noise limits.
The planets at largest separation are detected for the closest stars,
so it is straightforward to show that $dN_{det}/d\zeta \propto
\zeta^{-(3/b+1)} = \zeta^{[-4,-5/2,-7]}$ in the Zodi, PSF and EZ limits,
which is indeed the slope of the lines at large $\zeta$.  Since the
EZ flux limit grows most slowly with radius, the observed slope
will generally be $\propto \zeta^{-7}$.

\section{Maximizing over exposure time and survey volume} \label{maxtv}

I assume for the time being that the telescope observes a single type
of star at a single wavelength.  Then, if fewer stars are observed
out to a distance $D_s < D_{max}$ (i.e. ${\cal N} < 1$), more time can be spent on each
star making it possible to detect more planets since these stars are
closer.  Furthermore, it may pay to vary the exposure time with the
distance to the star, say $T_{exp} \propto D^{-\kappa}$.   Well, it
turns out that choosing $\kappa=0$ and ${\cal N}=1$ is close to optimal.
By increasing $\kappa$ to $\sim$0.5 (i.e. spending more time on closer stars)
the total number of planets found can be increased by only $\sim$0.5\%
compared to the $\kappa=0$ case (for either $\delta= 1/6$ or $1/8$).
The reason is that the most planets are
detected near $D_s$ due to the larger volume at larger distance,
but because these stars are more numerous, they also primarily
determine what the exposure time should be, so the total planet sensitivity
at that distance is roughly constant.  By making $\kappa$ slightly negative, 
then you can detect slightly more planets around closer stars, but this is 
counterbalanced by the fewer planets detected around more distant stars, and 
thus the total number of planets detected remains nearly constant.
One advantage of spending more time on closer stars is that there will
be more detections at higher signal-to-noise.  Specifically, the number
of stars with $S/N$ above $(S/N)_{det}$ scales as 
$(S/N)_{det}^{-1/(1+\delta\kappa)}$.  Since this has a weak dependence on 
$\kappa$, the gain is not significant, so for the rest of the paper I assume 
a uniform exposure time with distance, i.e.  $\kappa=0$.

\begin{figure}
\centerline{\psfig{file=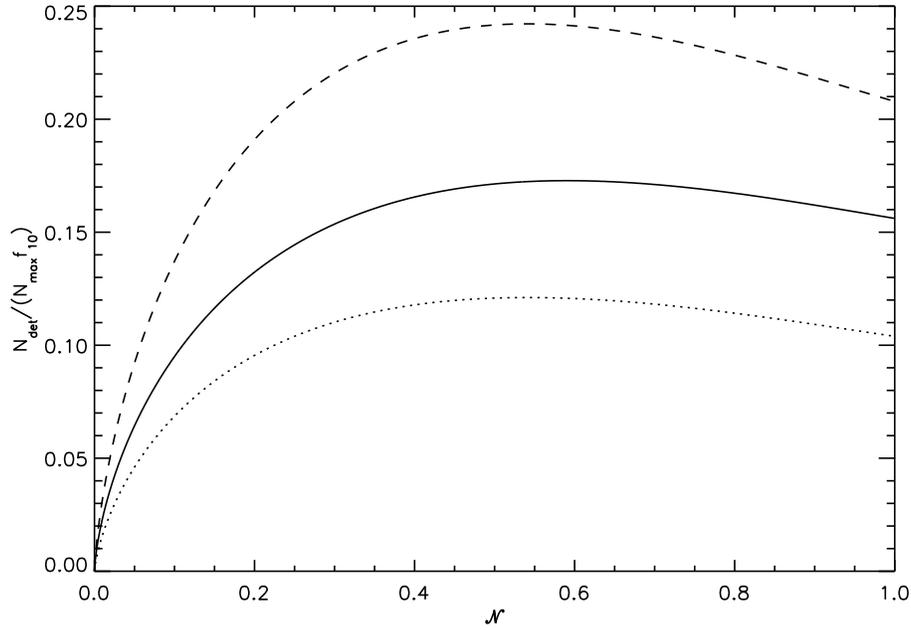,width=5in}} %Figure 4
\caption{The dependence of the number of detected planets, $N_{det}$
on the number of stars surveyed, ${\cal N}$, both in terms of $N_{max}$.
Solid line is for the Zodi limit, dashed line is for the Exo-zodi limit, 
and dotted line is the PSF limit.}
\label{fig04}
\end{figure}

A reduction in the number of stars can lead to an improvement in the number
of planets detected as more time is spent observing closer stars.  Figure
\ref{fig04} shows the dependence of $N_{det}$ on ${\cal N}$ computed from
\begin{equation}
{N_{det}\over N_{max}} = 3{\cal N}^{-3\delta} \int_0^{{\cal N}^{1/3+\delta}}d\eta \eta^2 E(\eta)
\end{equation}  
for each of the three noise limits.
For
the zodi-limited case, I find ${\cal N}=0.59$ optimizes the total number
of planets detected with an increase
in the number of planets by $\sim$11\% over the value at ${\cal N}=1$, while 
in the Exo-zodi and PSF limited
cases, ${\cal N}=0.54$ is optimum with an increase in the number of detected
planets by $\sim$16\% over the value at ${\cal N}=1$.  Even though only about 
half the number of stars are surveyed, the planet detection fraction roughly doubles 
causing a larger number of planets detected.  In terms of distance, this corresponds 
to observing stars out to $D_s \sim 0.75 D_{max}$ rather than out to $D_{max}$.

For the general case where all three sources of noise contribute, one must
optimize
\begin{equation}
N_{det}({\cal N},{\cal D}_1,{\cal D}_2) = \left(\Omega_s n_* D_{max}^3(1)\right) {D_{max}^3({\cal N}) \over D_{max}^3(1)}
\int_0^{{\cal N}^{1/3}D_{max}(1)/D_{max}({\cal N})} d\eta \eta^2 E(\eta,\eta_{EZ},\eta_{PSF}),
\end{equation}
where I have evaluated ${\cal D}_1,{\cal D}_2$ at ${\cal N}=1$, i.e., at 
$T_{exp}=T_s/N_{max}$.  The first term in parentheses is equal to $3N_{max}$.
The Zodi-noise dominated case is for ${\cal D}_1={\cal D}_2 = 0$ or
$\eta_Z=1, \eta_{EZ}=0, \eta_{PSF}=0$, while
the Exo-zodi noise dominated case is for ${\cal D}_1=\infty$, ${\cal D}_2=0$
or $\eta_Z=0, \eta_{EZ}=1, \eta_{PSF}=0$, and
the PSF-noise dominated case is for ${\cal D}_1=0$ and ${\cal D}_2 = \infty$
or $\eta_Z=0, \eta_{EZ}=0, \eta_{PSF}=1$.  The quantities $(\eta_{EZ},\eta_{PSF})$ can
be expressed as a function of $[{\cal N},{\cal D}_1,{\cal D}_2]$ with
the relations
\begin{eqnarray} \label{etaezpsfeqn}
\eta_{EZ}^6 &=& y^3 {\cal D}_1^6,\cr
\eta_{PSF}^6 &=& y^3 {\cal D}_2^6,\cr
y &=& (1-\eta_{EZ}^6 -\eta_{PSF}^6)^{1/4}\cr
y^4+{\cal R}y^3-{\cal N}^{-1} &=&0,\cr
{\cal R}&=&{\cal D}_1^6 +{\cal D}_2^6.\cr
\end{eqnarray}
Coincidentally, the equation for $y$ is same as the equation for
$(D_{max}/D_Z)^2$ (equation \ref{eqndmax}). 
For a grid of values of $[{\cal D}_1,{\cal D}_2]$ I solve
these equations for $\eta_{EZ},\eta_{PSF}$ as a function of ${\cal N}$ 
and then numerically compute 
$E(\eta,\eta_{EZ},\eta_{PSF})$ and $N_{det}[{\cal N},{\cal D}_1,{\cal D}_2]$.

I have carried out this numeric integration and then maximized $N_{det}$
with respect to ${\cal N}$.  I find that the optimum values of ${\cal N}$
vary from 0.54-0.60 for the entire range of ${\cal D}_{1,2}$, with a mean
of 0.57 and standard deviation of 0.02.  So,
in optimizing surveys I will just use the mean value of ${\cal N}=0.57$
since this only leads to a small error of at most 5\%.  The optimum
value of $N_{det}$ varies by a factor of 2 (between the PSF and Exo-zodi
limits, equation \ref{deteff}), so I have fit an equation to
the numerical results:
\begin{equation} \label{ndeteqn}
{N_{det} \over N_{max}} \approx f_{10} 
a_1 \left[\ln{\left(\eta_{EZ}^{12}+4a_0^{-4}
(1-\eta_{EZ}^6+\eta_{PSF}^6(a_0^{-2}-1)\right)^{1/2}-\eta_{EZ}^6
\over 2(1-\eta_{EZ}^6+\eta_{PSF}^6(a_0^{-2}-1))}\right]^{1/2},
\end{equation}
where $a_0=0.4249$ and $a_1=0.1287$.  This equation ranges from 12\%$f_{10}$ 
(for PSF-dominated noise with $\eta_{PSF}=1$) to 24\%$f_{10}$ (for Exo-zodi
dominated noise with $\eta_{EZ}=0$) and it agrees with the numerically
computed values of $N_{det}$ within $<1$\% fractional error!  The number
of stars being surveyed is $N_s={\cal N}N_{max}$, so the planet detection
fraction of stars surveyed is actually [29\%,44\%,21\%]$f_{10}$ in the 
[Zodi,Exo-zodi,PSF] cases; in practice the detection fraction will range between
these values.

\section{Optimizing the choice of stars} \label{rankstars}

Up until now, I have assumed that all stars are identical:  same radius, 
same spectral shape, same luminosity, same planet frequency.  However, surveys 
will be carried out for stars with a wide range of properties, so the question 
is:  which are the best to observe?

The answer to this should be obvious:  prioritize the stars which have
the highest probability for detection of planets.  
For a particular survey telescope, planet type, and exposure time,
$E(\eta,{\cal D}_1,{\cal D}_2)$ should be computed for each star and then stars can 
be ranked in decreasing order of $E$.  If one particular source of noise is expected
to dominate for all stars, then computation of $E$ can be sidestepped and stars
can be ranked just based on the survey properties.

To do this, I define a new parameter 
\begin{equation}
\beta  \propto \eta^{-1/\delta},
\end{equation}
that depends solely on the properties of the stars at the observational wavelength used
in computing $\eta$,
\begin{equation} \label{betadef}
\beta \equiv
\begin{cases} L_\nu^2 L_{\odot,\nu}^{-1} D^{-8} \tau_Z^{-1} & \text{Zodi-limit} \cr
 L_\nu D^{-6} \tau_{EZ}^{-1} & \text{Exo-zodi limit}\cr
 L_\nu D^{-6} & \text{PSF-limit}
\end{cases}
\end{equation}
Then, to choose stars I can simply rank stars from largest $\beta$
to smallest $\beta$ and observe in this order to maximize the number of
detected planets.  The parameter $\beta$ has the advantage that it can be defined
independent of the parameters of any particular survey so that the ranked
list of stars simply depends on the wavelength of observation and dominant
noise.

For now, I will assume that the planet frequency is independent of stellar 
type and that stars are chosen of the same metallicity.  For a given stellar type 
of uniform density one can show that $N(>\beta) = N_\beta \beta^{-3\delta}$.  
The normalization, $N_\beta$, is computed from
\begin{equation} \label{nbeta}
N_\beta \equiv
{\Omega_s \over 3}\int dM {dn \over dM} \begin{cases} L_\nu^{3/4}L_{\odot,\nu}^{-3/8} 
\int d\tau_Z p(\tau_Z) \tau_Z^{-3/8} & \text{Zodi-limit} \cr
L_\nu^{1/2} \int d\tau_{EZ} p(\tau_{EZ}) \tau_{EZ}^{-1/2} & \text{Exo-zodi limit}\cr
L_\nu^{1/2} & \text{PSF-limit},
\end{cases}
\end{equation}
where $p(\tau_Z)d\tau_Z$ is the probability distributions of $\tau_Z$ for the local zodiacal 
light; similarly, $p(\tau_{EZ})d\tau_{EZ}$ is the probability distribution for exo-zodiacal
light (both assumed to be independent of stellar spectral type).
I measure $L_\nu$ in units of $4\pi$ Jy pc$^2$ and $D$ in units of pc.

\begin{figure}
\centerline{\hbox{\psfig{figure=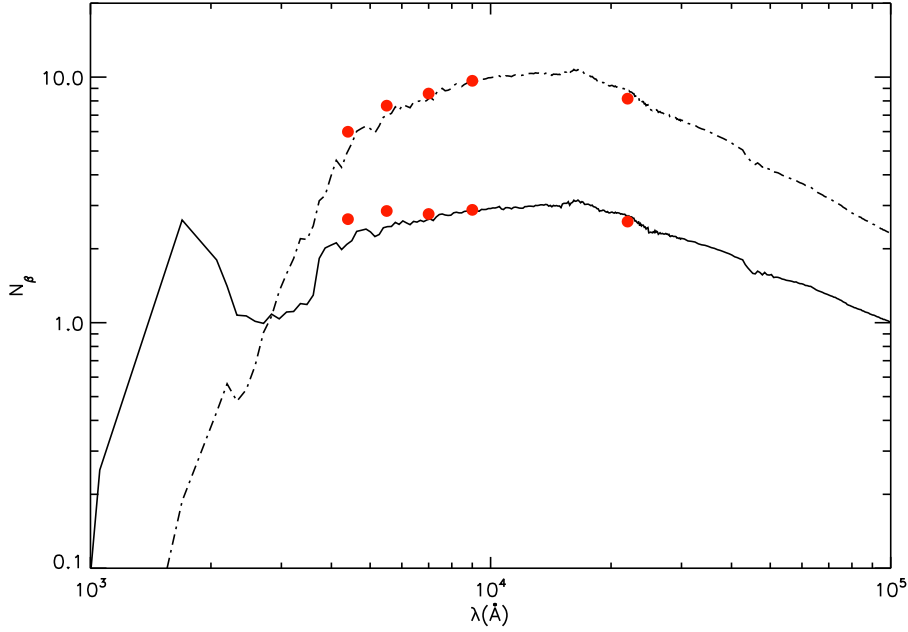,width=5in}}}
\caption{Normalization of $\beta$ distribution as a function of wavelength
for main sequence stars with $\Omega_s=4\pi$,  solid line is Zodi-limit while 
dashed-dot is Exo-zodi 
and PSF limits.  The large dots are computed using the $B,V,R,I$ 
and $K$ band luminosity functions of nearby main sequence stars.  For
the Zodi and Exo-zodi limits, $\tau_{Z,EZ}$ has been set to one.  The
units of $N_\beta$ are chosen so that the specific luminosity has units of
$4\pi$ Jy pc$^2$ and distance has units of pc.}
\label{fig05b}
\end{figure}

For computing $N_\beta$ I use the present-day disc mass function of main
sequence stars given by \citet{rei05}, the mass-temperature and 
radius-temperature
relations given by \citet{har88}, and stellar atmosphere models given by 
\citet{hau99}.  My computed $N_\beta(\lambda)$ is
shown in Figure \ref{fig05b} where it is compared with a model of the disc 
luminosity function of dwarf stars in $B, V, R, I$ and $K$ \citep{jar94}.
I have checked that this result agrees with the sample of stars from 
\citet{all99} (which is only complete for $T_{eff} > 6000$ K, so I have 
only compared in this range of temperatures).
About half of $N_\beta$ is due to
low temperature stars ($T_{eff} < 5000 K$) in the $I$-band, while more than 
one third is due to these stars in $B$ and $V$.  This indicates that
late-type stars should be included in coronagraphic planet searches due
to their greater number and thus greater proximity.  This function can be 
used to compute the optimum wavelength of observation as shown below. 

With $N_\beta$ in hand, the maximum number of stars to survey is
\begin{equation} \label{nmaxall}
N_{max}=\begin{cases}
N_Z=\left(N_\beta^{8/3}{\cal G} D_{tel}^8T_s[1 AU]^2 \lambda^{-6}\Theta_{IWA}^{-4}\right)^{3/11} & \text{Zodi-limit}\cr
N_{EZ}=\left(N_\beta^2{\cal G} D_{tel}^6 T_s \lambda^{-4} \Theta_{IWA}^{-2}\right)^{1/3} & \text{Exo-zodi limit}\cr
N_{PSF}=\left(N_\beta^2{\cal G} D_{tel}^6 T_s \lambda^{-2} C_0^{-1} \lambda_0^{-2}\Theta_{IWA}^{-4}\right)^{1/3} & \text{PSF limit},
\end{cases}
\end{equation}
where ${\cal G}$ is defined in equation \ref{calgdef}.
As before, if all three noise sources contribute then $N_{max}$ can be computed
from equation \ref{nmaxeqn} since ${\cal D}_1,{\cal D}_2$ are weakly dependent
on spectral type.
Now, as in the case of a survey of stars uniform stellar type, the number of detected
planets can be increased slightly if the number of stars surveyed is ${\cal N}=N_s/N_{max}=$54-59\%.
So, the cutoff is determined by
\begin{equation} \label{eqn44}
\beta_s = \left[{N_\beta \over {\cal N} N_{max}}\right]^{1/(3\delta)}.
\end{equation}
So, all stars with $\beta > \beta_s$ can be chosen to optimize the survey.
This guarantees that all stellar types will have a cutoff in distance at
the same fraction of $D_{max}$.
Since I am assuming a constant exposure time per star, then the time will
be divided up between various spectral types according to the integrand
in equation \ref{nbeta}.  For example, in the PSF-limited case, stars
of luminosity $L_\nu$ will have time spent $dT_s \propto (dn/dM)L_\nu^{1/2}dM$.

\section{Optimizing the wavelength of observation} \label{optlambda}

Since $N_{det}$ is a function of wavelength, we need to compute the
wavelength dependence of $N_{det}/N_{max}$ and $N_{max}$.
To compute the first quantity, ${\cal D}_1$ and ${\cal D}_2$ should be computed from 
equation \ref{caldeqn} for G-type stars (which are most common) at a
range of wavelengths.  Since
${\cal D}_{1,2}$ are weakly dependent on spectral type, these can be used
for optimizing a survey of all stellar types.  Then, $\eta_{EZ}$ and
$\eta_{PSF}$ can be computed from equations \ref{etaezpsfeqn} by computing $R$ 
from the fifth equation, solving 
for $y$ from the fourth equation (a quartic), and then 
plugging $y$ into the first two equations.  Then, equations
\ref{nmaxapprox} and \ref{nmaxall} can be used to compute $N_{max}$ from the other telescope,
stellar and survey parameters, while  equation \ref{ndeteqn} can be used to
compute the number of detected planets from $\eta_{EZ,PSF}$ and $N_{max}$.
This procedure yields $N_{det}(\lambda)$, which I carry out for 
a range of satellites discussed in the next section.

\begin{figure}
\centerline{\hbox{\psfig{figure=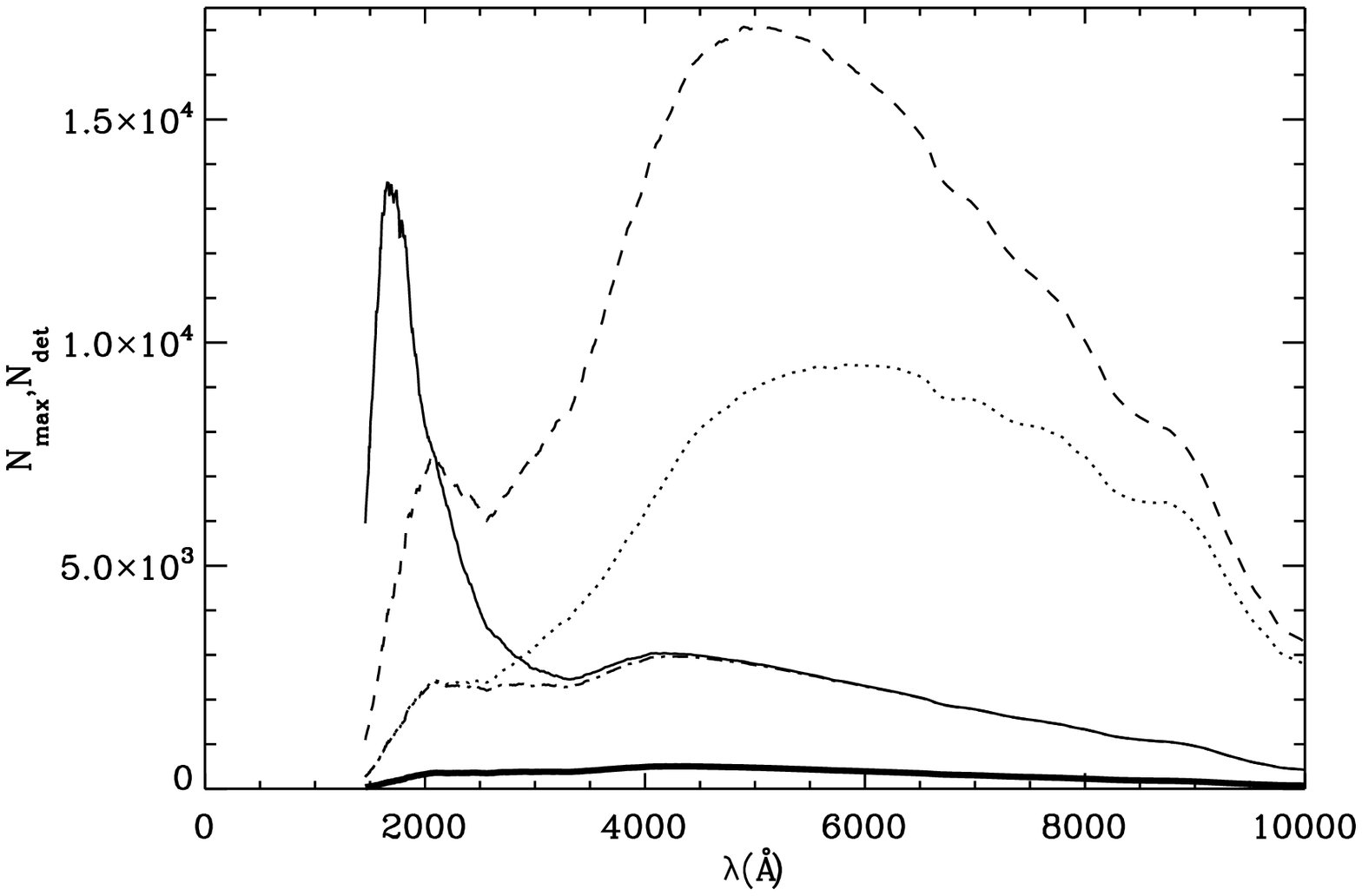,width=5in}}}
\centerline{\hbox{\psfig{figure=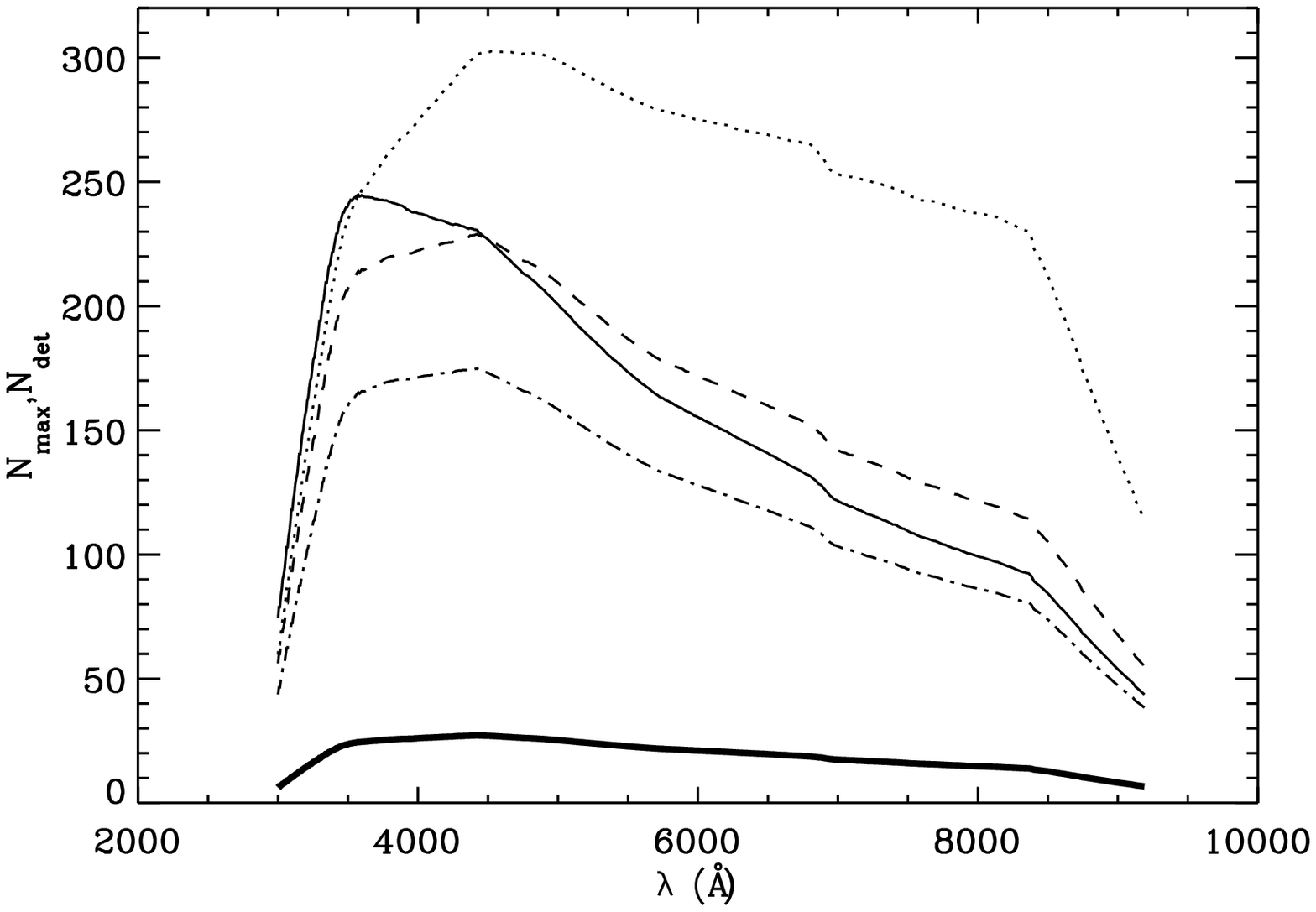,width=5in}}}
\caption{The maximum number of survey stars, $N_{max}$, and number of
detected planets, $N_{det}$, as a function of 
wavelength for a population
of planets with identical albedos to Jupiter (top) and Earth (bottom).
The solid line is $N_{max}$ for the Zodi-limited noise case, the dashed line is
the Exo-zodi limit, and the dotted line is the PSF-limit.  The dash-dot
line is $N_{max}$ from the combined noise limits for a survey with parameters 
of the {\it TOPS} telescope surveying all main-sequence stars (top panel) and 
parameters of {\it TPF-C} surveying only G dwarfs (bottom panel).
The dark solid lines are the number of detected planets, $N_{det}$, for an 
optimized survey with $f_{10}=1$.}\label{fig05c}
\end{figure}

The general wavelength scaling can be derived if one of the three noise limits
dominates since $N_{max}$ with wavelength as
\begin{equation}
N_{max} \propto \begin{cases}
\left(N_\beta^{8/3} \epsilon \lambda^{-6} p_\lambda^2\right)^{3/11} & \text{Zodi limit} \cr
\left(N_\beta^2 \epsilon \lambda^{-4} p_\lambda^2\right)^{1/3} & \text{Exo-zodi limit} \cr
\left(N_\beta^2 \epsilon \lambda^{-2} p_\lambda^2\right)^{1/3} & \text{PSF limit} \cr
\end{cases}
\end{equation}
assuming $\Delta \ln{\lambda}=$ constant.
I plot $N_{max}$ versus wavelength for Earth and Jupiter albedos 
shown in Figure \ref{fig05c}.
The albedo of the Earth I take from observations
of earthshine by \citet{woo02} and at shorter wavelengths from
direct observations of the Earth with GOME \citep{bur99}.  The albedo
of Jupiter I take from \citet{kar94} in the optical and
\citet{wag85} in the ultraviolet.  I have assumed a 20\% bandpass
for computing these curves and I assume that $\epsilon_\lambda$ is
independent of wavelength.

Figure \ref{fig05b} shows that the peak of $N_\beta$ is near $1.6$ micron
(this is due to the large number of stars cooler than the Sun).
However, due to the $\lambda^{-[6,4,2]}$
term, the peak in the number of detected planets is at a {\it much}
shorter wavelength, $\lambda_J=[3961,4785,5826]$\AA\  
for Jovian albedo planets and $\lambda_\oplus = [3953,4615,4794]$\AA\
for Earth albedo planets in the [Zodi,Ezo-zodi,PSF] noise limits,
respectively.  %Both curves have very broad regions which
For Earth-like planets a sharp drop occurs
shortward of 3000~\AA\ due to the absorption opacity of ozone.
Thus, surveys in the $B$ and $V$ bands should have the 
highest efficiency in detecting planets.

\section{Comparing proposed satellites} \label{comptel}

With the analytic relations derived in the previous sections in hand, I can now
compare various proposed coronagraphic planet imaging satellites.
A survey of the
SPIE conference proceedings ``Future EUV/UV and Visible Space 
Astrophysics Missions and Instrumentation'' and ``High-Contrast 
Imaging for Exo-Planet Detection" turns up at least six planned
coronagraphic satellites:  {\it EPIC/OPD} \citep{men03}, {\it ECLIPSE}
\citep{tra03,hul03}, {\it TPF-C} \citep{bro03,tra06}, {\it ESPI} \citep{lyo03},
{\it UMBRAS} \citep{sch03}, {\it ExPO} \citep{gez03}, and {\it TOPS}
\citep{guy06}.  For comparison, I also include the
{\it Hubble Space Telescope} \citep{bro90} and the {\it Hubble Space 
Telescope} with a corrective secondary 
mirror, {\it HST$^*$} as proposed by \citet{mal95}.
I also include the ``{\it DREAM}" satellite which is my version of
what an ideal design for {\it TPF-C} would be capable of.
   The parameters of these different missions reported in these
citations are listed in Table \ref{tab00}.

To compare these various satellites, I define a figure of merit which
I call ``planet imaging power", or PIP, \citep[analogous to the ``A-$\Omega$"
\'etendue defined for large surveys, ][]{cla04,kai04}.
It is given simply by 
\begin{equation} \label{pip}
PIP = \begin{cases}
(D_{tel}^{24}\epsilon^{3} \Theta_{IWA}^{-12})^{1/11} & \text{ Zodi limit},\cr
(D_{tel}^{18}\epsilon^{3} \Theta_{IWA}^{-6})^{1/9} & \text{ Exo-Zodi limit},\cr
(D_{tel}^{18} \epsilon^{3} \Theta_{IWA}^{-12} C_0^{-3})^{1/9}  & \text{ PSF limit}.
\end{cases}
\end{equation} 
The number of planets that may be detected with a given telescope is
simply proportional to $PIP$ assuming that a particular source of noise
dominates.   I have computed these for each proposed satellite in Table \ref{tab00},
which shows that TPF-C is the top-ranked proposed satellite in all three
noise limits.

To compare these satellites, I assume a throughput of $\epsilon=$3\%
(independent of wavelength - although for {\it TOPS} and {\it DREAM}
I assume an efficiency of 20\%), a mission duration of $T_s=3$yr, a solid 
angle of $\Omega_s=4\pi$, and a planet albedo identical to either the Earth 
or Jupiter.  I compare four surveys (1) a 3-year survey of solely G-type
(Sun-like) stars optimized for either jupiters or earths; (2) a 3-year
survey of all nearby main-sequence stars optimized for either jupiters
or earths.  Additional assumed parameters 
are $\Delta \ln{\lambda}=0.2$, $(S/N)_{det}=10$, $\tau_Z= 6.8\times 10^{-10}$
sr$^{-1}$
(which is the weighted value in the anti-solar direction),
$\tau_{EZ}=2.7\times 10^{-9}$ sr$^{-1}$ (which assumes an edge-on zodiacal 
cloud like the Sun's around every star), a quasi-Lambert
phase function, and, for the G-star survey, I assume a stellar
number density of $n_*=5\times 10^{-3}$ pc$^{-3}$.
For Jupiter-albedo planets I assume $R_p=7.15\times 10^9$ cm and 
and for Earth-albedo planets I assume $R_p=6.38\times 10^8$ cm.
I compute the optimum $T_{exp}$, $N_{max}$, $N_{det}$, and $f_{det}=N_{det}/(N_s f_{10})$
for each proposed survey, listed in Tables \ref{tab01}-\ref{tab04} ($max$ is 
abbreviated $m$ for some quantities in the table).
Two examples for the G-dwarf survey of earth-sized planets with 
{\it TPF} and a survey of  all main sequence stars for jupiter-sized
planets with {\it TOPS} are shown in Figure \ref{fig05c}.
For the computed exposure times, I also compute $D_{max}$ and $r_{max}$
for Sun-like stars.  I assume that
there is one planet per decade of radius for {\it every} star
(that is, $f_{10}=1$).
Since the current data favors $f_{10} \sim 10$\%
for Jovian planets (Tables \ref{tab01} and \ref{tab03}), these numbers should be reduced by a 
factor of $\sim$ $f_{10}$.  

Since each telescope has slightly different design parameters,
the values in Tables \ref{tab01}-\ref{tab04} are only an
approximate guide.  However, a few points are striking.   
All of the detection efficiencies are between 21-32\% of the
average number of stars per decade of radius for the surveyed
stars.
Although the {\it ECLIPSE} and {\it EPIC} telescopes 
have similar sizes, the {\it EPIC} telescope is much more
powerful due to its smaller inner working angle.   For finding
terrestrial planets about G dwarfs, the {\it TPF-C} is the only
one with a statistically significant chance due to its large
aperture size. More
recent designs for {\it TPF-C} include a
larger mirror than I have used here, so this number may be
an underestimate.  With a 4m mirror, an optimized survey
of G dwarfs with {\it TPF-C} is most sensitive to planets at $\sim$2 AU, outside the 
habitable zone (if fewer stars are surveyed, then TPF-C can be optimized for 
the habitable zone, but many fewer planets will be detected).
Jupiter-sized planets are optimized for detection at a few AU, but
will be detected at a large range of radii.
Comparing tables \ref{tab02} and \ref{tab04} shows that there is a factor of $\sim$3-5 increase in
the number of detected earth-sized planets if {\it all} stars surveyed
rather than just G-dwarfs.  It is remarkable that the much smaller
1.2m {\it TOPS} telescope will have a significant chance of detecting a
few Earth-sized planets if $f_{10}$ is not too small.
The optimum wavelengths of detection range from 4000-4900 \AA\ for
earth-albedo planets and 4400-5900 \AA\ for Jupiter-albedo planets;
thus, the $B$ and $V$ bands are optimal for planet detection.
The last two columns give the ratios of the distance maxima in the
different noise limits.  For all of the surveys the different sources
of noise contribute at a similar level - this is not surprising as
the telescopes were designed with the realization that it does not
pay to reduce the PSF wings much below the Zodi and Exo-zodi limits.
This is apparent in the both panels of Figure
\ref{fig05c} where the Zodi and PSF noise contribute equally for
the TOPS survey of jupiters, while Zodi and Exo-zodi noise contribute
equally for the TPF-C survey of earths around G-dwarfs.  

\begin{table}
  \caption{Proposed satellite parameters.}\label{tab00}
  \begin{tabular}{lcccccc}
  \hline
   Satellite &  $\Theta_{IWA}$ & $D_{tel}$ & $C_0$ & $PIP_Z$ & $PIP_{EZ}$ & $PIP_{PSF}$ \cr
             &                 &  (m)      &@550nm \cr
\hline
{\it ECLIPSE}&  4.0 &  1.8 &   1.e-09  &  0.3 &  0.4 &  0.16 \cr
{\it EPIC}   &  1.5 &  1.5 &   1.e-09  &  0.6 &  0.5 &  0.41 \cr
{\it ESPI}   &  4.0 &  1.5 &   3.e-07  &  0.2 &  0.3 &  0.02 \cr
{\it ExPO}   &  2.9 &  3.0 &   1.e-09  &  1.3 &  1.4 &  0.68 \cr
{\it TPF-C}  &  3.9 &  4.0 &   5.e-11  &  1.8 &  2.0 &  2.20 \cr
{\it UMBRAS} &  3.5 &  1.0 &   1.e-08  &  0.1 &  0.1 &  0.03 \cr
{\it HST}    &  3.0 &  2.4 &   1.e-06  &  0.8 &  0.9 &  0.04 \cr
{\it HST$^*$}&  3.0 &  2.4 &   5.e-10  &  0.8 &  0.9 &  0.52 \cr
  {\it TOPS} &  1.5 &  1.2 &   1.e-10  &  0.6 &  0.6 &  1.06 \cr
  {\it DREAM}&  1.5 &  8.0 &   5.e-11  & 38.7 & 28.6 & 59.17 \cr
\hline
\hline
\end{tabular}
\end{table}
 
\begin{table}
  \caption{Expected number of detectable Jovian planets surveying all stars.}\label{tab01}
  \begin{tabular}{lccccccccc}
  \hline
  Satellite & $T_{exp}$ & $N_s$ & $N_{det}/f_{10}$ & $f_{det}$ & $\lambda_{m}$ & $D_{m}$ & $r_{m}$ & ${\cal D}_1$ & ${\cal D}_2$\\
  Acronym   & (hr)      &           &                  &     (\%)  &   (\AA)         &    (pc)         &   (AU)          &                         & \\
\hline
Equation(s) &  \ref{texpeqn} &  \ref{calneqn},\ref{nmaxeqn},\ref{nmaxall}& \ref{ndeteqn},\ref{etaezpsfeqn} && &\ref{dmaxeqn} & \ref{rmaxeqn} & \ref{caldeqn},\ref{dmaxzodi},\ref{dmaxez} & \ref{caldeqn},\ref{dmaxzodi},\ref{dmaxpsf}\\
\hline
{\it ECLIPSE}& 41.7 &    630 &   152 & 24.2 &   5189.3 &  23.6 &   6.5 &  0.69 &  0.94\cr
{\it EPIC}   & 18.9 &   1392 &   366 & 26.4 &   4897.5 &  29.7 &   3.5 &  0.85 &  0.86\cr
{\it ESPI}   &311.8 &     84 &    17 & 20.9 &   5848.7 &  12.3 &   4.6 &  0.62 &  2.11\cr
{\it ExPO}   &  9.7 &   2710 &   662 & 24.4 &   5189.3 &  38.0 &   4.5 &  0.77 &  0.95\cr
{\it TPF-C}  &  5.5 &   4790 &  1368 & 28.6 &   4412.9 &  44.6 &   4.6 &  0.79 &  0.68\cr
{\it UMBRAS} &198.9 &    132 &    28 & 21.7 &   5584.0 &  14.2 &   6.6 &  0.65 &  1.22\cr
{\it HST}    &123.9 &    212 &    44 & 20.9 &   5848.7 &  16.7 &   2.9 &  0.68 &  2.56\cr
{\it HST$^*$}& 14.5 &   1810 &   465 & 25.7 &   4897.5 &  33.0 &   4.8 &  0.76 &  0.87\cr
  {\it TOPS} & 15.5 &   1691 &   503 & 29.8 &   4278.7 &  31.0 &   3.9 &  0.82 &  0.59\cr
  {\it DREAM}&  0.3 & 105044 & 32520 & 31.0 &   4412.9 & 117.9 &   2.3 &  0.97 &  0.61\cr
\hline
\hline
\end{tabular}
\end{table}

\begin{table}
  \caption{Expected number of detectable Jovian planets around Sun-like stars.}\label{tab03}
  \begin{tabular}{lccccccccc}
  \hline
  Satellite & $T_{exp}$ & $N_s$ & $N_{det}/f_{10}$ & $f_{det}$ & $\lambda_{m}$ & $D_{m}$ & $r_{m}$ & ${\cal D}_1$ & ${\cal D}_2$\\
  Acronym   & (hr)      &           &                  &     (\%)  &   (\AA)         &    (pc)         &   (AU)          &                         & \\
\hline
Equation(s) &  \ref{texpeqn} &  \ref{calneqn},\ref{numstar},\ref{nmaxall}& \ref{ndeteqn},\ref{etaezpsfeqn} && &\ref{dmaxeqn} & \ref{rmaxeqn} & \ref{caldeqn},\ref{dmaxzodi},\ref{dmaxez} & \ref{caldeqn},\ref{dmaxzodi},\ref{dmaxpsf}\\
\hline
{\it ECLIPSE}&130.8 &   200 &    49 & 24.8 &   5149.4 &  27.7 &   7.6 &  0.66 &  0.90\cr
{\it EPIC}   & 63.5 &   413 &   110 & 26.8 &   4897.5 &  35.3 &   4.1 &  0.81 &  0.81\cr
{\it ESPI}   &911.4 &    28 &     6 & 20.9 &   5562.5 &  14.7 &   5.2 &  0.60 &  2.07\cr
{\it ExPO}   & 30.9 &   850 &   211 & 24.9 &   5149.4 &  44.9 &   5.3 &  0.74 &  0.91\cr
{\it TPF-C}  & 19.0 &  1382 &   399 & 28.9 &   4568.8 &  52.6 &   5.6 &  0.74 &  0.64\cr
{\it UMBRAS} &592.4 &    44 &     9 & 21.9 &   5562.5 &  16.9 &   7.8 &  0.62 &  1.17\cr
{\it HST}    &362.1 &    72 &    15 & 20.9 &   5562.5 &  20.1 &   3.3 &  0.65 &  2.51\cr
{\it HST$^*$}& 47.3 &   555 &   145 & 26.2 &   4897.5 &  38.9 &   5.7 &  0.73 &  0.82\cr
  {\it TOPS} & 55.2 &   476 &   141 & 29.8 &   4533.7 &  36.9 &   5.0 &  0.77 &  0.54\cr
  {\it DREAM}&  1.0 & 27282 &  8365 & 30.7 &   4568.8 & 142.8 &   2.9 &  0.90 &  0.57\cr
\hline
\hline
\end{tabular}
\end{table}

\begin{table}
  \caption{Expected number of detectable Earth-like planets surveying all stars.}\label{tab02}
  \begin{tabular}{lccccccccc}
  \hline
  Satellite & $T_{exp}$ & $N_s$ & $N_{det}/f_{10}$ & $f_{det}$ & $\lambda_{m}$ & $D_{m}$ & $r_{m}$ & ${\cal D}_1$ & ${\cal D}_2$\\
  Acronym   & (day)      &           &                  &     (\%)  &   (\AA)         &    (pc)         &   (AU)          &                         & \\
\hline
Equation(s) &  \ref{texpeqn} &  \ref{calneqn},\ref{nmaxeqn},\ref{nmaxall}& \ref{ndeteqn},\ref{etaezpsfeqn} && &\ref{dmaxeqn} & \ref{rmaxeqn} & \ref{caldeqn},\ref{dmaxzodi},\ref{dmaxez} & \ref{caldeqn},\ref{dmaxzodi},\ref{dmaxpsf}\\
\hline
{\it ECLIPSE}& 32.1 &    34 &     7 & 22.0 &   4903.2 &   9.0 &   2.3 &  0.91 &  1.27\cr
{\it EPIC}   & 13.1 &    83 &    20 & 24.8 &   4881.3 &  11.3 &   1.3 &  1.11 &  1.12\cr
{\it ESPI}   &299.6 &     3 &     0 & 20.9 &   4903.2 &   4.4 &   1.4 &  0.83 &  2.99\cr
{\it ExPO}   &  7.5 &   145 &    32 & 22.5 &   4903.2 &  14.4 &   1.6 &  1.02 &  1.28\cr
{\it TPF-C}  &  3.0 &   370 &   101 & 27.4 &   4412.6 &  18.2 &   1.9 &  1.02 &  0.89\cr
{\it UMBRAS} &182.2 &     6 &     1 & 21.0 &   4903.2 &   5.2 &   2.1 &  0.86 &  1.69\cr
{\it HST}    &119.1 &     9 &     1 & 20.9 &   4903.2 &   6.0 &   0.9 &  0.90 &  3.62\cr
{\it HST$^*$}& 10.2 &   107 &    25 & 23.6 &   4881.3 &  12.8 &   1.9 &  1.00 &  1.13\cr
  {\it TOPS} &  7.5 &   145 &    43 & 30.0 &   4079.4 &  12.5 &   1.5 &  1.09 &  0.79\cr
  {\it DREAM}&  0.1 &  8634 &  2768 & 32.1 &   4181.3 &  44.4 &   0.8 &  1.28 &  0.83\cr
\hline
\hline
\end{tabular}
\end{table}

\begin{table}
  \caption{Expected number of detectable Earth-like planets surveying Sun-like stars.}\label{tab04}
  \begin{tabular}{lccccccccc}
  \hline
  Satellite & $T_{exp}$ & $N_s$ & $N_{det}/f_{10}$ & $f_{det}$ & $\lambda_{m}$ & $D_{m}$ & $r_{m}$ & ${\cal D}_1$ & ${\cal D}_2$\\
  Acronym   & (day)      &           &                  &     (\%)  &   (\AA)         &    (pc)         &   (AU)          &                         & \\
\hline
Equation(s) &  \ref{texpeqn} &  \ref{calneqn},\ref{numstar},\ref{nmaxall}& \ref{ndeteqn},\ref{etaezpsfeqn} && &\ref{dmaxeqn} & \ref{rmaxeqn} & \ref{caldeqn},\ref{dmaxzodi},\ref{dmaxez} & \ref{caldeqn},\ref{dmaxzodi},\ref{dmaxpsf}\\
\hline
{\it ECLIPSE}& 94.4 &    11 &     2 & 21.7 &   4522.8 &  10.0 &   2.4 &  0.96 &  1.38\cr
{\it EPIC}   & 43.6 &    25 &     6 & 24.2 &   4422.5 &  13.0 &   1.4 &  1.18 &  1.23\cr
{\it ESPI}   &848.6 &     1 &     0 & 20.9 &   4533.0 &   4.8 &   1.4 &  0.88 &  3.26\cr
{\it ExPO}   & 22.7 &    48 &    10 & 22.2 &   4522.8 &  16.1 &   1.7 &  1.08 &  1.39\cr
{\it TPF-C}  & 11.0 &    99 &    27 & 27.3 &   4422.5 &  20.8 &   2.1 &  1.06 &  0.93\cr
{\it UMBRAS} &518.8 &     2 &     0 & 21.0 &   4533.0 &   5.6 &   2.1 &  0.92 &  1.85\cr
{\it HST}    &337.4 &     3 &     0 & 20.9 &   4533.0 &   6.5 &   0.9 &  0.96 &  3.94\cr
{\it HST$^*$}& 31.7 &    34 &     7 & 23.0 &   4422.5 &  14.5 &   1.9 &  1.06 &  1.24\cr
  {\it TOPS} & 31.8 &    34 &    10 & 30.3 &   4422.5 &  14.6 &   1.9 &  1.10 &  0.78\cr
  {\it DREAM}&  0.6 &  1718 &   556 & 32.4 &   4422.5 &  52.9 &   1.0 &  1.30 &  0.82\cr
\hline
\hline
\end{tabular}
\end{table}

\section{Optimizing the metallicity of survey stars} \label{metals}

Since it is known that planets of period less than 4 years show
a frequency proportional to the square of the metallicity of the host star 
\citep{fis05,san04}, the question arises as to how to distribute observing time 
among stars of different metallicities.  Although coronagraphic surveys 
for giant planets may tend to find planets at periods longer than a few
years, it may be 
reasonable to assume that the planet frequency-metallicity correlation will 
also hold for the distribution of planets discovered in a coronagraphic 
survey.  I will assume that the exposure 
time is adjusted in proportion to the expected number of planets
per star raised to some power $\gamma$, $ T_{exp}(X) \propto f_{10}^\gamma$
($X=$[Fe/H]).
If I assume that the magnitude of
a given type of star is independent of the metallicity, then 
the number of planets found about that stellar type is
\begin{equation}
N_{det} \propto {\int f_{10}^{1+3\gamma\delta/(1+3\delta)} {dn_* \over dX} dX
\over \left(\int f_{10}^{\gamma} {dn_* \over dX} dX\right)^{3\delta/(1+3\delta)}}.
\end{equation}
By differentiating with respect $\gamma$, the number of planets
is maximized for $\gamma=1+3\delta$, meaning that each $X$ bin should
be weighted by metallicity to the $\sim$third power.  Using the metallicity
distribution measured by \citet{val05} I find that the number
of planets can be increased by 14\% or 19\% for $\gamma=1+3\delta$ 
(with $\delta=1/8$ or $1/6$) versus $\gamma=0$ (i.e. no preference for higher 
metallicity stars).

\section{Discussion and Conclusions} \label{conclusions}

Since detection of planets is the first-order goal of a coronagraphic survey,
maximizing $N_{det}$ has some interesting implications:

1) {\it Survey duration:}  The total number of planets detected
scales as $N_{det}\propto T_s^{3/11}$ in the Zodi limit or $\propto T_s^{1/3}$ in the
Exo-zodi and PSF limits, which means that about ten times as much time 
must be spent to double the number of detected planets.

3) {\it Telescope design optimization:} The number of planets detected scales 
as $N_{det} \propto PIP$ (given in equation \ref{pip}) in the three different noise limits.
The greatest gains come from first increasing the telescope aperture,
second decreasing the inner working angle, and third decreasing the PSF
wing contrast (if the PSF limits the detection) or increasing the telescope 
efficiency.  %I have only assumed 

4) {\it Optimum wavelength:}  This must be computed by combining the optimum
number of stars to survey in all three noise limits.  For Earth albedo the peak
tends to lie in the B band, while for Jupiter albedo it lies in the V band
(however, different telescope and survey parameters might shift the optimum
wavelength to another region).  In all cases the optimum wavelength is
shortward of the peak of the stellar photon number flux.

5) {\it Stellar spectral types:}   The number of planets detected scales
with spectral type as $N_{det} \propto dn_*^{8/11}L_\nu^{6/11}$ in the
Zodi noise dominated limit, while $N_{det} \propto dn_*^{2/3}L_\nu^{1/3}$ in
the Exo-zodi and PSF dominated limits.  Stars of G type dominate the
number of detected planets (assuming that the frequency of planets is
independent of stellar type).  If main sequence stars of all types are
surveyed, then the number of planets can be increased by a factor
of $\sim 3$ over surveys that only target G-type stars.

6) {\it Planet size:}  
I have computed the number of detected planets assuming a single
planet size (equation \ref{planetdist}).  However, we know that there
will be a distribution of sizes of planets, in which case the number of 
planets detected at other sizes scales as ${df\over dR_p} R_p^{[3/2,2]}$ 
for smaller planets (in the [Zodi,Exo-zodi/PSF] limits), while 
for larger planets it has a more complicated scaling.
So, for example, Earth-sized planets would need to be $\sim$10$^2$ 
times more abundant than giants to be detected in similar quantities 
by a survey optimized for giants.

7) {\it Signal-to-noise:}  Since the number of planets in an
optimized survey above a signal-to-noise $(S/N)_0$ scales as
the volume within which those planets can be detected, then 
\begin{equation}
N_{det}[S/N > (S/N)_0]  = N_{det}[S/N > (S/N)_{det}] \begin{cases}
\left({(S/N)_{det} \over (S/N)_0}\right)^{3/4} & \text{Zodi-limit,} \cr
{(S/N)_{det} \over (S/N)_0} & \text{Exo-zodi/PSF limits.} \end{cases}
\end{equation}

8) {\it Planet-star distance:}  Since $D_{max}$ sets the
maximum distance to which planets may be detected, $r_{max}=
\theta_{IWA}D_{max}/\sin{\alpha_{max}}$ sets the planet-star separation at which 
the bulk of the planets will be detected (most planets are
detected close to $r_{max}$ since this is where the volume
of the survey is maximized).  Since G type stars dominate the
number of detections, $r_{max}$ should be computed  for G stars in 
all three noise limits, and the minimum value should provide a good 
estimate of the peak of the planet-star distance distribution.
If there is a reason to believe that the target planet population
is not distributed uniformly in $log(r)$, then the optimization of
the observing strategy will change only slightly.  I have run some
test cases with a planet-star separation distribution that is
a power law with radius, and I find that the optimum ${\cal N}$
is still around 0.6.  The reason for this is that the volume
term dominates - at large distances the number of stars surveyed grows
rapidly, so it does not pay to greatly decrease the number of surveyed
stars.   However, the values of $N_{det}$ can change significantly
for a non log-uniform distribution of planet-star separations.  For
a planet-star separation that scales as $d f/dr \propto r^{-(1+\xi)}$,
total number of planets detected will scale as $r_{max}^{-\xi}$, and
so the specific value of $r_{max}$ for a given telescope and survey
strategy will determine how many planets are detected.

If one is concerned with targeting habitable-zone planets, then
a different optimization strategy is required (which will likely involve
observing more nearby stars); I will defer a study of this to future 
work \citep[see ][ for a numerical approach]{bro05}.

9) {\it Fraction of detectable planets:}  I computed that if all
stars are surveyed within $D_{max}$, then $\sim 21-44$\% of
the number of planets per decade in radius will be detected (even
though a smaller number of stars are surveyed than in an unoptimized
survey); the satellites modeled here show a range of 21-32\%.  %, while
Multiple visits might increase this yield as
discussed by \citet{bro04a}, which likely requires a Monte-Carlo
approach, so I defer a study of this issue to future work.

10) {\it Metallicity:}  If the planet frequency increases
as the square of metallicity (as it does for the known
extrasolar giant planets), then the exposure time should be
scaled as the $\sim$cube of the metallicity.  This increases the
number of detected planets by about 15-20\% over the number detected
with a uniform exposure time.

11) {\it Proposed telescopes:}  The {\it TPF-C} is by far
the most powerful and ambitious of the proposed coronagraphic imaging 
telescopes.  However, a much smaller telescope {\it TOPS} has
a significant chance of detecting Earth-size/albedo planets if
they are not too rare.  The main competitive strengths of
{\it TOPS} are (a) smaller inner working angle and (b) high
throughput.  Thus, a modified version of {\it TPF-C} with a
larger mirror, higher throughput, and smaller inner working angle
(I dub ``{\it DREAM}") would allow more than an order of magnitude 
increase in the number of detected Earth-like planets and would
have a peak sensitivity in the habitable zone.

Finally, I wish to comment on the few assumptions I have made
which should be studied in future work. I have assumed that the 
albedo and phase function is independent of the separation between 
planet and star and spectral type of the host star, and that the 
phase function is independent of wavelength.
The simulated Jupiter-mass planetary atmospheres in \citet{sud05}
show an inverse square-law dependence with radius in the V-band from
$\sim 2-20$ AU, show a nearly constant phase function at $> 45^\circ$
from 0.2-4 AU, and show a geometric albedo that differs by at
most 0.2 in the optical over 2-15 AU from the albedo at 6 AU.
In regions that differ from an inverse square law, the results
in this paper will change quantitatively, but probably not qualitatively.
The Lambert phase function shape is a very poor description of
cloudy atmospheres which can have a peak in the brightness at
crescent phase \citep{sud05};  however, this effect is more significant at
longer wavelengths which are unfavorable due to poorer angular
resolution.

I have assumed that the PSF has a hard ``edge" at the inner
working angle within which no planet can be detected, and that
the wings of the PSF have a constant contrast.  This is also
clearly an oversimplification, but a full study will require
knowing precisely the properties of a given instrument.  I have
assumed that the telescope is circular, while plans for TPF-C indicate
that an oblong telescope may give better detection properties without
much added cost.  My analysis can accommodate an
oblong mirror which has an inner working angle set by the long axis,
which can be taken to be $2R_{tel}$.  Then, the area of the telescope
is smaller than $\pi R_{tel}^2$ by the ratio of the axis lengths
which can be included in the efficiency factor, $\epsilon$.
Observations at several roll angles can be used to circularize
the inner working angle of an oblong telescope which also leads to 
a hit in the efficiency by the inverse of the number of roll angles.

I have assumed that the telescope efficiency is 
independent of wavelength.  This might be achieved, for example, by 
observations with a superconducting tunneling junction (STJ) array 
\citep{pea98}, but until then one must repeat my analysis
with the efficiency of a particular telescope folded in.
STJ detectors have the additional advantage that they measure the
energy of each photon allowing detection of broad spectral features
and subtraction of speckles which have different widths in each
waveband.   I have also assumed that there is no limit on exposure
time - it may be that pointing control, cosmic rays, or other factors
will limit the exposure time per star, which has not been taken
into account in this analysis.

I have assumed that the planet frequency is independent of
spectral type; however, enough data to quantify the
variation of planet frequency with spectral type currently
do not exist, although some predictions have been made
\citep{ida04,lau04}.    One of the most significant uncertainties
in these calculations is the strength of the Exo-zodiacal light,
of which there are currently no observational constraints.  
It is likely that the Exo-zodiacal dust distribution will depend
on the age, spectral type, planet and minor planet distributions
around given stars, so there is no simple way to predict or
parameterize this source of noise.  Further studies in the
mid-infrared would be very helpful in constraining this factor
and assessing how important it will be in limiting planet
detection.

I have ignored the effect of $\theta_{OWA}$ since it only causes 
obscuration of planets orbiting the nearest stars, blocking a fraction
$\sim(\theta_{IWA}/\theta_{OWA})^2$ which is typically quite a small
number.  I have primarily concentrated on a constant exposure time
for all stars, but weighting the exposure time towards nearer stars
can cause a slight increase in the signal-to-noise for each detection
and increase the number of nearby planets detected at close distances
to the host star around stars closer to the Sun without decreasing the 
number of detected planets significantly.

I have assumed for the blind surveys that there is no prior information
about the existence of planets about these stars.  However, given
that of order one thousand stars have been surveyed to date with the
radial velocity technique, it is likely that some parameter space
for planets can be excluded for some survey stars.  Currently the
surveys are complete to about $K \sim 15$ m/s for long period planets,
$\sim 3$ years \citep{cum04}, while the peak of the surveys for Giant planets is
about 3-15 years, so the current constraints will likely rule out only
a small region of parameter space.  However, as radial velocity surveys 
improve their sensitivity and time baseline these constraints will
become more stringent.  Incorporating this information into target selection 
and exposure time will be left for future work.

If you wish to use the codes used to compute the figures or tables, please
visit the author's web page at the University of Washington Astronomy Department.

\section{Appendix}

To make the notation in this paper easier to navigate, here are
two tables (\ref{tab06},\ref{tab07}) referencing the symbols used throughout, the units of each
quantity, and the section or equation the first time each symbol appears in the paper.

\begin{table} 
  \caption{Definition of Roman symbols.}\label{tab06}
  \begin{tabular}{llcc}
  \hline
Symbol & Usage & Units & Equation \cr
  \hline
$a_0,a_1$  & Numbers used in approximation to $N_{det}/N_{max}$ & -- & \ref{ndeteqn} \cr
$C(\lambda)$& Intensity contrast of PSF & -- & \S \ref{csa}\cr
$C_0$& Intensity contrast of PSF at a fiducial wavelength & -- & \S \ref{csa}\cr
$D$        & Distance of star from the Earth & cm or pc & \ref{eqn01} \cr
${\cal D}_1$& Ratio of $D_Z$ to $D_{EZ}$ & -- &\ref{caldeqn}\cr
${\cal D}_2$& Ratio of $D_Z$ to $D_{PSF}$ & -- &\ref{caldeqn}\cr
$D_{max}$  & Maximum distance a planet can be detected & cm or pc & \S \ref{sss} \cr
$D_s$      & Maximum distance of survey & cm or pc & \ref{maxnumstars}\cr
$D_Z$  & $D_{max}$ in the Zodi noise dominated limit & cm or pc & \ref{dmaxzodi} \cr
$D_{EZ}$  & $D_{max}$ in the Exo-zodi noise dominated limit & cm or pc & \ref{dmaxez} \cr
$D_{PSF}$  & $D_{max}$ in the PSF noise dominated limit & cm or pc & \ref{dmaxpsf} \cr
$D$        & Distance of star from the Earth & cm or pc & \ref{eqn01} \cr
$E(\eta)$  & Expection value of the number of planets that can be detected & -- & \ref{ndetint} \cr
$\tilde E$ & Approximation to $E(\eta)$ & -- & \ref{expsection} \cr
$f_{10}$   & Expected number of planet per decade of $r$ & -- & \ref{planetdist} \cr
$f_{det}$   & Fraction of stars observed with detected planets, per decade radius & -- & \S \ref{comptel} \cr
$G$        & Auxiliary quantity used in defining $D_{max}$ & -- & \ref{dmaxeqn} \cr
${\cal G}$ & Quantity used in computing $D_{max}$ and $N_{max}$ & cm$^4$ erg$^{-1}$ s$^{-1}$  &  \ref{calgdef} \cr
$h$        & Planck's constant & erg s & \ref{eqn07} \cr
$H$        & Heaviside step function    & -- & \ref{eqn01} \cr
$L_\nu$    & Star's specific luminosity & erg s$^{-1}$ Hz$^{-1}$ & \ref{eqn07} \cr
$L_{\nu,\odot}$    & Sun's specific luminosity & erg s$^{-1}$ Hz$^{-1}$ & \ref{eqn08} \cr
$M$        & Stellar mass & $M_\odot$ & \ref{eqn01}\cr
${\cal N}$ & Ratio of $N_s$ to $N_{max}$ & -- & \ref{calneqn} \cr
$N_{det}$  & Number of detected planets & -- & \ref{eqn01} \cr
$N_{max}$  & Maximum number of stars that can be surveyed & -- & \ref{numstar},\ref{nmaxeqn}\cr
$N_{Z,EZ,PSF}$  & Maximum number of stars that can be surveyed in 3 noise limits & -- & \ref{numstar},\ref{nmaxall}\cr
$N_s$      & Number of stars in survey  & -- & \ref{texpeqn} \cr
$N_\beta$  & Normalization of $\beta$ distribution & various & \ref{nbeta}\cr
$n_*$      & Number density of stars of a particular spectral type & pc$^{-3}$ & \S \ref{maxnumstars}\cr
$p_\lambda$& Geometric albedo of planet as a function of wavelength & -- & \ref{eqn03}\cr
$PIP$      & Planet imaging power (in each noise limit) & various & \ref{pip}\cr
$Q_{EZ}$   & Variance of noise from exo-zodiacal light & -- & \ref{sneqn},\ref{eqn09} \cr
$Q_{Z}$   & Variance of noise from Zodiacal light & -- & \ref{sneqn},\ref{eqn08} \cr
$Q_{PSF}$   & Variance of noise from stellar PSF & -- & \ref{sneqn} \cr
$Q_{B}$   & Variance of noise from other backgrounds & -- & \ref{sneqn} \cr
$Q_p$      & Number of photons detected from planet in an exposure & -- & \ref{eqn03} \cr
$Q_*$      & Number of photons detected from star in an exposure & -- & \ref{eqn03} \cr
$r$        & Planet-star separation     & cm or AU & \ref{eqn01} \cr
$r_{1}$  & Minimum $r$ at which $\theta=\theta_{IWA}$  & cm or AU & \ref{eqn12}\cr
$r_{2}$  & Maximum $r$ at which $S/N = S/N_{det}$ & cm or AU & \ref{eqn32}\cr
$r_{2,Z}$  & Maximum $r$ at which $S/N = S/N_{det}$ in Zodi noise limit & cm or AU & \ref{rmaxzodi}\cr
$r_{2,EZ}$  & Maximum $r$ at which $S/N = S/N_{det}$ in Exo-zodi noise limit  & cm or AU & \ref{ezsnr}\cr
$r_{2,PSF}$  & Maximum $r$ at which $S/N = S/N_{det}$ in PSF noise limit  & cm or AU & \ref{eqrmax}\cr
$r_{max}$ & Planet-star separation at maximum elongation & cm or AU &  \ref{rmaxeqn}\cr
${\cal R}$ & Auxiliary quantity used in defining $\eta_{EZ}$, $\eta_{PSF}$ & -- & \ref{etaezpsfeqn}\cr
$R_p$      & Planet radius              & cm & \ref{eqn01} \cr
$R_{tel}$  & Radius of telescope        & cm & \ref{eqn07} \cr
$s$        & Quantity used in defining $E(\eta)$ & -- & \ref{eqleqn} \cr
$S/N$      & Signal-to-noise ratio      & -- & \ref{eqn01} \cr
$S/N_0$      & Fiducial signal-to-noise ratio      & -- & \S \ref{conclusions} \cr
$S/N_{det}$& $S/N$ required for detection & -- &\ref{eqn01} \cr
$S_{fac}$  & PSF sharpness & -- & \ref{eqn08} \cr
$T_{exp}$  & Exposure time per star & sec & \S \ref{csa} \cr
$T_s$      & Total time of survey & sec & \S \ref{csa}\cr
$V$        & Auxiliary quantity used in defining $D_{max}$ & -- & \ref{dmaxeqn} \cr
$w$        & Auxiliary quantity used in defining $dN_{det}/d\zeta$ in PSF noise limit& -- & \ref{eqn35}\cr
$x_{\pm}$  & Quantity used in defining $E(\eta)$ & -- & \ref{eqleqn}\cr
$X$        & $[Fe/H]$ log of metallicity & dex & \ref{eqn01} \cr
\hline
\end{tabular}
\end{table}
\begin{table} 
  \caption{Definition of Greek symbols.}\label{tab07}
  \begin{tabular}{llcc}
  \hline
Symbol & Usage & Units & Equation \cr
  \hline
$\alpha$   & Planet phase angle         & rad & \ref{eqn01} \cr
$\alpha_\pm$& Phase angle where resolution and flux limits cross & rad & \ref{eqn33} \cr
$\alpha_{max}$& Phase angle where planet-star sky angle is maximum & rad & \S \ref{zodinoise} \cr
$\beta$ & Parameter used to rank priority of observing different stars in each noise limit & various & \ref{betadef}\cr
$\beta_s$ & Value of $\beta$ at $D_s$ & -- & \ref{eqn44}\cr
$\gamma$  & Exponent of exposure time versus planet frequency as a function of metallicity & -- & \ref{metals}\cr
$\Gamma$   & Quantity used in defining $E(\eta)$ & -- & \ref{eqleqn}\cr
$\delta(x)$   & Dirac delta function       & --  & \ref{planetdist} \cr
$\delta$   & Exponent of $T_{exp}$ in equations for $D_{max}$ & --  & \ref{dmaxzodi} \cr
$\Delta$   & Quantity used in defining $E(\eta)$ & -- & \ref{eqleqn}\cr
$\Delta \ln{\lambda}$ & Fractional bandwidth of telescope & -- & \S \ref{csa} \cr
$\epsilon$ & Total throughput           & --  & \S \ref{csa} \cr
$\zeta$    & Ratio of $r$ to $r_{max}$ & -- & \S \ref{ddd}\cr
$\eta$     & Ratio of $D$ to $D_{max}$ & -- & \S \ref{ddd}\cr
$\eta_Z$   & Ratio of $D_{max}$ to $D_Z$ & -- & \ref{eqn30} \cr
$\eta_{EZ}$   & Ratio of $D_{max}$ to $D_{EZ}$ & -- & \ref{eqn30} \cr
$\eta_{PSF}$   & Ratio of $D_{max}$ to $D_{PSF}$ & -- & \ref{eqn30} \cr
$\theta$   & Sky angle separation of planet \& star & rad & \ref{eqn01} \cr
$\theta_{IWA}$ & Inner working angle of coronagraph & rad &\ref{eqn01} \cr
$\Theta_{IWA}$ & Dimensionless inner working angle & -- & \ref{eqn04} \cr
$\kappa$   & Power law exponent for exposure time as a function of distance & -- & \ref{maxtv}\cr
$\lambda$  & Wavelength    & \AA & \ref{eqn03} \cr
$\lambda_0$  & Reference wavelength for $C(\lambda)$   & -- & \S \ref{csa} \cr
$\Sigma$        & Auxiliary quantity used in defining $dN_{det}/d\zeta$ in PSF noise limit& -- & \ref{eqn35}\cr
$\xi$      & Power law exponent of $df/d\ln{r}$ versus $r$ & -- & \S \ref{csa} \cr
$\tau_Z$       & Relates solar flux to zodiacal surface brightness & sr$^{-1}$ & \ref{eqn08} \cr
$\tau_{EZ}$       & Relates stellar flux at tangent to line of sight to exo-zodiacal surface brightness & sr$^{-1}$ & \ref{eqn09} \cr
$\phi$     & Quantity used in defining $E(\eta)$ & -- &\ref{eqleqn} \cr
$\Phi(\alpha)$ & Planet phase function & -- & \ref{eqn03} \cr
$\Omega_s$ & Solid angle of planet survey & rad & \ref{eqn01} \cr
\hline
\end{tabular}
\end{table}

\section{acknowledgements}

I would like to thank Eric Ford for carefully reading the
manuscript and for comments which improved the paper.  The referee,
Scott Gaudi, also gave comments that made the paper more complete and
improved the presentation.

\bibliography{MF1545rv2}
\bibliographystyle{mn2e}
\clearpage

\end{document}